\documentclass[lettersize,journal]{IEEEtran}
\usepackage{amsmath,amsfonts}
\usepackage{algorithmic}
\usepackage{algorithm}
\usepackage{array}
\usepackage[caption=false]{subfig}
\usepackage{textcomp}
\usepackage{stfloats}
\usepackage{url}
\usepackage{verbatim}
\usepackage{graphicx}
\usepackage[hidelinks]{hyperref}
\usepackage{cite}
\usepackage{booktabs}
\usepackage{multirow}
\usepackage{siunitx}
\usepackage{tabularx}
\usepackage{makecell}
\usepackage{enumitem}
\usepackage{orcidlink}
\usepackage[normalem]{ulem}
\usepackage{tabularx}
\usepackage{arydshln}
\usepackage{array}

\begin{document}

\title{RF-Zero-Wire: Design and Analysis of Multi-Hop Low-latency Symbol-synchronous RF Communication
}
\author{Xinlei Liu\,\orcidlink{0009-0007-0789-0041}, Andrey Belogaev\,\orcidlink{0000-0003-2609-399X}, Jonathan Oostvogels\,\orcidlink{0000-0003-4554-5445}, Bingwu Fang\,\orcidlink{0000-0002-1173-8014}, Danny Hughes\,\orcidlink{0000-0002-0750-3693}, Jeroen~Famaey\,\orcidlink{0000-0002-3587-1354}~\IEEEmembership{Senior Member,~IEEE}
\thanks{This work was supported in part by the European Union’s Horizon Europe Framework under Grant 101093046 - project OpenSwarm, and in part by the Flemish Government under FWO Project G019722N - LOCUSTS. Jonathan Oostvogels is funded by the Research Foundation - Flanders~(FWO), grant number 11H7923N. The computing resources and services for this work were supported by the HPC core facility CalcUA of the University of Antwerp, and Flemish Supercomputer Center~(VSC), funded by the Flemish government. 

Xinlei Liu, Andrey Belogaev, and Jeroen Famaey are with IDLab, University of Antwerp - imec, Belgium.~(email: \{Xinlei.Liu, Andrey.Belogaev, Jeroen.Famaey\}@uantwerpen.be)

Jonathan Oostvogels, Bingwu Fang, and Danny Hughes are with DistriNet, KU Leuven, Belgium.~(email: \{Jonathan.Oostvogels, Bingwu.Fang, Danny.Hughes\}@kuleuven.be)
}}

\IEEEpubid{\begin{minipage}{\textwidth}\centering\ \\[12pt] 
  Copyright (c) 2026 IEEE. Personal use of this material is permitted.\\ 
  However, permission to use this material for any other purposes must be obtained from the IEEE by sending a request to pubs-permissions@ieee.org.
\end{minipage}}




\maketitle

\begin{abstract}
The latency gap between wired and wireless networks poses a challenge in the adoption of wireless technologies in latency-sensitive scenarios. The gap is especially notable in multi-hop communication typical for industrial sensor networks and robotic swarms. The main reason behind it is that commonly used wireless protocols rely on store-and-forward routing and costly overhead procedures to avoid interference. This article introduces RF-Zero-Wire, an RF-based symbol-synchronous communication protocol. Instead of relaying the whole frame per hop in a store-and-forward manner, nodes concurrently relay the frame symbol by symbol, without the need for tight time synchronization. Based on data collected in real-world experiments, we reveal that the inevitable carrier frequency offsets~(CFOs) introduced by imperfect crystal oscillators cause a beating effect under concurrent symbol transmissions. This is characterized by periodic constructive and destructive interference, which significantly affects reliability. Subsequently, a thorough simulation study shows how the beating problem can be overcome with error correction codes. RF-Zero-Wire allows achieving an end-to-end latency of less than \SI{1}{\milli\second} for a small $4$-byte frame transmitted across $5$ hops. Moreover, latency is shown to increase only by \SI{0.16}{\percent} per extra hop for $16$-byte frames, which is negligible compared to the over \SI{100}{\percent} per-hop latency increase observed in store-and-forward protocols. The trade-offs between network reliability and CFO range, communication distance, node density, and achievable data rate are studied in large-scale experiments based on simulation. 
\end{abstract}

\begin{IEEEkeywords}
IoT, WSN, robotic swarms, symbol-synchronous communication, carrier frequency offset.
\end{IEEEkeywords}

\section{Introduction}
\label{sec1}
\IEEEPARstart{T}{he} Internet of Things~(IoT), embedding numerous physical devices into a network and allowing them to communicate with each other, has experienced remarkable development over the past two decades~\cite{8584051, Kopetz2022}. IoT has been widely applied in various domains, such as smart healthcare~\cite{li2024review}, industrial automation~\cite{7883994}, agriculture~\cite{XU202210}, and transportation~\cite{9125435}. As a significant part of IoT, wireless sensor networks (WSNs), consisting of sensor-equipped nodes, are widely adopted to monitor and transfer sensor readings like temperature, humidity, motion, or chemical gases~\cite {lee2015internet}.

Multi-hop communication is widely used by WSNs. It relies on store-and-forward relay techniques, where nodes repeatedly forward messages to neighboring nodes within a short range, with the goal of reaching a sink node connected to the wider network (e.g., Internet). It allows WSNs to be deployed in a large-scale area and extend network coverage with low expenditure on network infrastructure~\cite{barrachina2017multi}. In addition, robotic swarms, consisting of mobile robots that interact with each other and the environment to collaboratively complete various tasks (e.g., environment mapping, or resource gathering), share many characteristics with WSNs~\cite{NEDJAH2019100565,10639038}. 

\begin{figure*}[!t]
\centering
\subfloat[Store-and-forward routing]{\includegraphics[width=2.4in]{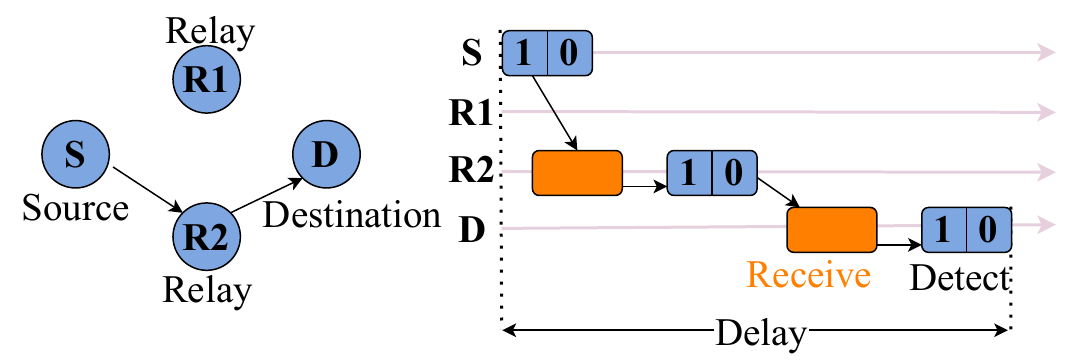}%
\label{sf}}
\hfil
\subfloat[Synchronous transmission]{\includegraphics[width=2.4in]{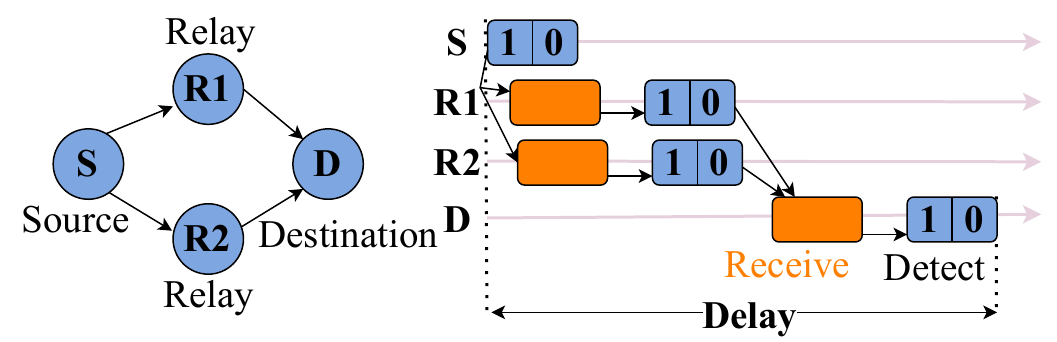}%
\label{ct}}
\hfil
\subfloat[Symbol-synchronous transmission]{\includegraphics[width=2.2in]{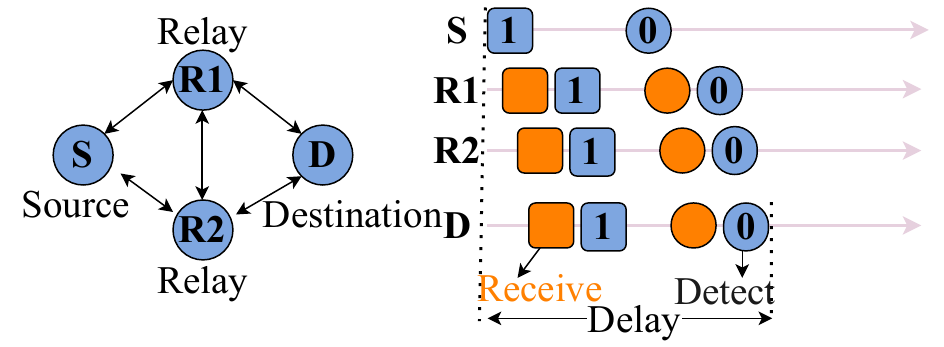}%
\label{ss}}
\caption{Multi-hop transmission flow for different methods, showing the transmission of two bits from a source $S$ across relays $R1$ and $R2$ to destination $D$}
\label{fig1}
\end{figure*}

Although WSNs have shown their distinctive strengths in many fields, existing WSNs are incapable of supporting critical applications, such as industrial control loops, robotic swarms, and vehicle-to-everything (V2X) networks, where there are strict latency, reliability, and power consumption requirements~\cite{8673568}. For instance, IoT for Industry 4.0 typically requires end-to-end latencies below \SI{1}{\milli\second} to broadcast the control decision and ensure safe and efficient operation, for example, in industrial control loops or safe human-cobot interactions~\cite{maia2024survey}. Low latency is also required for real-time interaction among multiple collaborative robots to ensure optimal cooperation in completing complex tasks~\cite{zheng2023integrated}.
\IEEEpubidadjcol
In addition, sensor information in vehicular networks needs to be delivered within sub-\SI{10}{\milli\second} for remote driving and emergency trajectory alignment~\cite{ficzere2023large}. Wireless multi-hop network protocols, such as 6TiSCH~\cite{thubert2021rfc}, ISA100.11a~\cite{petersen2011wirelesshart}, and WirelessHART~\cite{song2008wirelesshart} were proposed as low-latency and high-reliability solutions for industrial real-time control systems. However, the end-to-end latency of the network is still around hundreds of milliseconds over multiple hops~\cite{8368987}, and increases by tens of milliseconds with each additional hop~\cite{7536317}. Even with only one hop, the end-to-end latency is around tens of milliseconds, which is much higher than in widely used wired industrial control networks, like Controller Area Network~(CAN), whose latency is around hundreds of microseconds~\cite{davis2007controller}. The increased latency in wireless multi-hop networks is caused by their reliance on store-and-forward routing, combined with interference-avoiding channel access methods such as Time Division Multiple Access~(TDMA), Frequency Division Multiple Access~(FDMA), or Carrier Sense Multiple Access~(CSMA)~\cite {thubert2021rfc, petersen2011wirelesshart, song2008wirelesshart}. The store-and-forward routing principle, as shown in Fig.~\ref{sf}, requires the node to wait until complete reception of the frame before performing error detection, waiting for channel access, and finally relaying over the next hop. This generally results in high latency that increases by multiple packet transmission times with each hop. This is especially detrimental when transmitting longer frames. Also, many traditional protocols, depending on a predefined or semi-static routing tree, cannot easily be used in scenarios with mobile nodes, such as robot swarms.

The concept of synchronous transmissions has recently been proposed to make use of frame collisions instead of avoiding them~\cite{zimmerling2020synchronous}. It challenges the conventional view that frame collisions are always destructive. In contrast, it suggests that superimpositions of the same frames can potentially generate constructive interference at the receiver with high-precision time synchronization. This concept allows multiple relay nodes to concurrently relay the same frame without the need for expensive channel access procedures, as shown in Fig.~\ref{ct}. However, it still makes use of the store-and-forward principle and thus suffers from its drawbacks, i.e., the latency increases at least by one packet transmission time per hop. 

This article proposes a novel multi-hop communication protocol, evolving the concept of synchronous to symbol-synchronous transmission~(cf., Fig.~\ref{ss}), aiming to address the shortcomings of store-and-forward methods. The concept of symbol-synchronous transmission has recently been proposed by Oostvogels \textit{et al.} \cite{10.1145/3384419.3430897}. Their proposed Zero-Wire method was applied to wireless optical communications. Our aim is to adapt this concept to the RF domain, dubbed RF-Zero-Wire, to achieve a higher data rate and wider coverage. This is significantly more challenging, as overlapping signals in the optical domain are always constructive. In contrast, radio waves may interfere destructively due to phase and frequency offsets, especially since Zero-Wire does not assume tight time synchronization across relays. 

We consider the problem of information flooding, where the information needs to be transmitted to all nodes in the network from one initiator. It enables symbol-level concurrent transmissions, which means that each relay node does not need to wait for the reception of a whole frame before forwarding it to its neighbours. Compared with conventional synchronous transmissions based on frame-level concurrent transmissions, our protocol manages to reduce the end-to-end latency of multi-hop networks through the avoidance of the critical waiting time of frame reception. Moreover, high-precision time synchronization that is necessary in synchronous transmissions~\cite{5779066} is not needed in RF-Zero-Wire, further lowering the complexity of the deployment of the protocol. Although the concept is applicable to a wide range of frequency bands, we validate it in the $2.4$~GHz Industrial, Scientific, and Medical (ISM) band, due to its wide availability and frequent use for WSNs. In addition, we calibrate the key parameters related to the transceiver installed in each relay node of the network, through small-scale hardware experiments based on Adalm Pluto Software-Defined-Radios~(SDRs)\cite{adalm_pluto}. With the aid of these experiments, a realistic simulation is set up in MATLAB to evaluate network latency, reliability, and scalability. 

The main contributions of this article are as follows.
\begin{enumerate}
\item We present RF-Zero-Wire, an RF-based symbol-synchronous communication protocol that reduces the end-to-end network latency in dense wireless multi-hop networks, such as WSNs and robotic swarms. We define the modulation and demodulation procedures, as well as symbol-synchronous transmission strategies.
\item We study the challenges and limitations of deploying RF-based symbol-synchronous communication, based on small-scale hardware experiments, using SDRs. The effect of CFOs on symbol-level concurrent transmissions is specifically shown and analyzed. Our studies based on data from hardware experiments complement the previous simulation-based studies~\cite{liu2024low,10590570} and address their overly optimistic assumptions.
\item Based on hardware calibration, we develop a realistic MATLAB-based simulation framework to demonstrate that the proposed protocol can limit the latency below $1$~ms for small-frame transmissions~($4$-byte frames). In addition, we show that the latency scales well with the number of hops, paving the way for the deployment of large-scale distributed networks for latency-sensitive scenarios. Finally, we investigate how the network reliability depends on the system parameters.
\end{enumerate}

The remainder of this article is organized as follows. Section~\ref{sec2} reviews the related work for communication protocols in WSNs, with a specific focus on protocols based on concurrent transmissions. Section~\ref{sec3} provides details of the proposed symbol-synchronous RF-Zero-Wire protocol. Section~\ref{sec4} describes the hardware calibration experiments and shows some critical results, including the effect of carrier frequency offsets~(CFOs) on received signals of concurrent transmissions. Subsequently, Section~\ref{sec5} evaluates the performance of the proposed protocol using a large-scale simulation. Finally, Section~\ref{sec6} draws conclusions.

\section{Related work}
\label{sec2}

This section introduces some classical and efficient protocols that have been developed and applied in wireless multi-hop networks. Their limitations are analyzed, and the gaps addressed by our work are highlighted. 

\subsection{Multi-hop store-and-forward protocols}
For a reliable and low-latency wireless multi-hop network, numerous communication protocols prioritize internal collision avoidance as a primary design objective through carefully managing channel occupation and transmission timing. Some classical Media Access Control (MAC) techniques, TDMA, FDMA, and CSMA, avoid collisions by allocating unique timeslots, partitioning the spectrum, and carrier sensing, respectively~\cite{de2016ieee,hadded2015tdma,pan2020information,ziouva2002csma}. Although these MAC layer techniques improve communication reliability, they introduce hundreds to thousands of milliseconds of latency and tens of milliseconds of jitter~\cite{karaagac2018time, wang2016tdma, sgora2015survey}. However, latency-sensitive applications such as collaborative robotic swarms and industrial control loops require end-to-end latency and jitter in the order of milliseconds or less~\cite{10859271,8673568}. 

Several multi-hop network protocols for industrial WSNs have been proposed, building on a combination of TDMA and FDMA concepts, such as 6TiSCH~\cite{thubert2021rfc} and WirelessHART~\cite{song2008wirelesshart}. Despite optimized scheduling, the per-hop latency of 6TiSCH and WirelessHART remains in the order of $15$~ms, due to its complicated MAC layer design~\cite{7536317}. Zhang \textit{et al.}~\cite{9566795} propose a new scheduling method for multi-hop networks, where the relay node per hop is dynamically selected, rather than statically predefined as in WirelessHART and 6TiSCH. This method is inherently designed to reduce the part of the latency caused by scheduling. However, the latency introduced by store-and-forward is still not resolved.

\subsection{Concurrent transmission protocols}
Although the design details of 6TiSCH and WirelessHART are different, their relay mechanisms are similar, only allowing one node to relay the frames per hop to avoid collisions. Synchronous transmission challenges the view that frame collisions should be mitigated~\cite{zimmerling2020synchronous}, as the resulting interference is not always destructive. In contrast, under certain conditions, the superimposed frames from concurrent transmissions can increase signal strength, improving the probability of successful decoding and the interference resilience of the network~\cite{10129192}. Such synchronous transmission, strengthening the signals in the receiver, is particularly crucial for enabling a low-power wireless system, where multiple nodes are able to simultaneously transmit the same frames with lower power. Recently, a growing number of protocols have exploited synchronous transmissions. Glossy, a typical network based on synchronous transmission, leverages precise time synchronization to enable multiple nodes to broadcast the same frames simultaneously~\cite{5779066}. It utilizes the \mbox{IEEE 802.15.4} standard transceiver. Glossy indicates that if the time offset is less than \SI{0.5}{\micro\second}, concurrent transmissions based on the IEEE 802.15.4 standard frames will generate constructive interference, improving transmission reliability~\cite{liao2016revisiting}. Taking advantage of the Start Frame Delimiter~(SFD) in the \mbox{IEEE 802.15.4} frame, Glossy manages to synchronize the relay nodes within \SI{0.5}{\micro\second} to ensure constructive interference in the receiver. BlueFlood~\cite{nahas2021blueflood} applies the core ideas of Glossy based on precise time synchronization and synchronous transmission to Bluetooth communication. However, the differences between the physical layer of Bluetooth~5~\cite{sheikh2021adaptive} and IEEE 802.15.4 reduce the effectiveness and reliability of the approach. Chaos~\cite{landsiedel2013chaos} focuses on communication between multiple transmitters and receivers and supports data sharing over multiple nodes, which captures the strongest signals and neglects the weak signals.

\subsection{Beyond store-and-forward routing}
While the above-mentioned approaches have greatly contributed to improving network reliability and reducing energy consumption, they are not suitable for critical latency-sensitive applications, which are currently relying on wired networks due to their deterministic sub-ms latency and jitter. This is attributed to the fact that they all maintain the store-and-forward principle, which receives, inspects, and forwards the data on a per-frame basis. Specifically, in store-and-forward, each relay node stores the symbols and waits until all symbols in one frame are received and the frame checksum is verified. Subsequently, a scheduler tells the relay node how and when to forward the frame. Although a lot of efforts have been made to achieve fast routing and even eliminate routing~\cite{5779066,escobar2020using}, network end-to-end delays scale poorly with increasing hop count due to the nature of the store-and-forward principle. For instance, Glossy floods an \mbox{$8$-byte} frame within $3$ milliseconds, and one extra hop will add around \SI{500}{\micro\second} delay~\cite{5779066} in a typical configuration. However, Glossy requires very tight time synchronization within \SI{0.5}{\micro\second}, which is challenging to achieve for low-power and low-cost devices, where resource constraints and clock drift limit the feasibility of continuous fine-grained synchronization. 

If wireless multi-hop networks are to further decrease the latency below $1$~ms that simple wired networks maintain, the store-and-forward approach needs to be rethought. To this end, Oostvogels \textit{et al.} proposed Zero-Wire, a wireless symbol-synchronous transmission bus concept~\cite{10.1145/3384419.3430897} that solves the deficiencies of store-and-forward routing. In symbol-synchronous transmission, the processing object in each relay node is no longer a frame but a single symbol. Specifically, when the relay node successfully detects the symbol, it will forward it immediately to neighboring nodes instead of storing all symbols in a frame before relaying. Like synchronous transmissions, symbol-synchronous transmission still exploits the multiple concurrent transmissions to enable low-latency and low-power applications. With symbol-synchronous transmission, Zero-Wire deterministically lowered the end-to-end latency below $1$~ms for a \mbox{$2$-byte} frame over a \mbox{$25$-node} optical network. However, wireless optical communications are limited in their use cases, as they do not support long-distance transmissions and struggle in non-line-of-sight~(NLOS) scenarios. Zippy presents a bit-level network flooding method using RF transmissions~\cite{sutton2015zippy}. Like symbol-synchronous transmission, Zippy also broadcasts information symbol-by-symbol with an on-off keying~(OOK) transceiver. However, it encodes one symbol using repetition codes whose code length is equal to the maximum hop count over the network, resulting in an increase of code redundancy with the number of hops. Also, due to the redundant encoding in Zippy, its data rate is limited to $1.4$~kbps, and the end-to-end latency is tens of milliseconds for 2-byte packets.
\subsection{Emerging low-latency protocols}
Recently, some network protocols have emerged to decrease the end-to-end latency of multi-hop networks. The 6TiSCH protocol with a novel and optimized low-latency autonomous scheduling scheme is proposed for low-latency industrial IoT networks~\cite{pradhan20226tisch}. A single slot frame in this scheme is divided into several segments according to the number of hops, and each segment is only used for transmissions from nodes with identical hop counts. With this scheme, packets are forwarded hop by hop in consecutive segments, forming a pipeline-like flow toward the source node. Consequently, all nodes can send data packets to the source node in a single slot frame. The low end-to-end latency is achieved through reducing control-overhead delays, transmission collisions, and backoff and waiting times. However, this scheme requires tight time synchronization and is very sensitive to the number of hops in the network. Once the hop count changes, the segment structure needs to be reconfigured accordingly. The Multi-LoRa protocol enables low-latency multi-hop communication in a large-scale IoT deployment through optimizing the hardware architecture\cite{prade2022multi}. Each node in Multi-LoRa is equipped with two LoRa radios operating on different frequencies, which send different packet chunks at the same time, thus decreasing the end-to-end latency by cutting the packet transmission time. Although using two radios is more direct and simpler to decrease the latency, it introduces additional hardware costs and higher power consumption, which is inappropriate for low-power IoT devices. In addition, the multi-hop and multi-connection~\mbox{(MHMC)} network model with an optimized routing and traffic allocation strategy is proposed to reduce the end-to-end latency~\cite{ma2023routing}. Based on probability theory and statistical analysis, the authors derived the closed-form expression of end-to-end latency by modeling the transmission rate, subcarrier allocation, resource contention, retransmission attempts, and backoff delay. According to the derived mathematical expression, the hop counts, selection of relays, and resource allocations are respectively optimized for minimizing end-to-end latency. Nevertheless, the method highly relies on several probabilistic assumptions (e.g., traffic distribution and contention behavior), which may not be realistic in highly dynamic IoT environments. Table~\ref{table5} briefly lists the emerging low-latency protocols and corresponding latency, advantages, and limitations.

\begin{table*}[!t]
\begingroup
\caption{Emerging low-latency protocols for critical IoT applications\label{table5}}
\centering
\begin{tabularx}{\textwidth}{
    >{\raggedright\arraybackslash}p{2.2cm} | X | >{\centering\arraybackslash}p{2cm} | X | X
}
\hline
\textbf{Protocol} & \textbf{Key methods} & \textbf{Latency} & \textbf{Advantages} & \textbf{Limitations}\\
\hline
6TiSH with autonomous scheduling~\cite{pradhan20226tisch} &
Hop-based slotframe segment, pipeline-like packet forwarding &
\SI{300}{\milli\second} for $127$ bytes and $6$ hops &
Avoidance of centralized coordination and control exchanges &
Sensitive to hop counts\\
\hdashline
Multi-LoRa~\cite{prade2022multi} & Dual LoRa radio for multiplexing transmission & \SI{500}{\milli\second} for $32$ bytes and $6$ hops & Easier and more direct to implement & Higher radio cost and power consumption\\
\hdashline
MHMC~\cite{ma2023routing} & Mathematically modelling and minimizing the end-to-end latency based on probability theory & \SI{3}{\second} for $375$ bytes and $3$ hops & Closed-form latency formula enabling direct recalculation using other parameters & Several assumptions may not hold in practical IoT environments\\
\hdashline
Glossy~\cite{5779066} & Packet-synchronous and concurrent transmission & \SI{3}{\milli\second} for $8$ bytes and $8$ hops & Easy deployment on the commercial PHY layer of IEEE 802.15.4 & Highly dependent on a tight time synchronization of \SI{5}{\micro\second} \\
\hdashline
Zippy~\cite{sutton2015zippy} & Repetition codes and
sub-bit-synchronous and concurrent transmission & \SI{20}{\milli\second} for $1$ byte and $3$ hops & Exploit the carrier frequency randomization for the destructive interference of concurrent transmission & Very low data rate of $1.364$ kbps, limited by repetition codes\\
\hdashline
RF-Zero-Wire (our method) & Symbol-synchronous and concurrent transmission & Sub-\SI{1}{\milli\second} for $4$ bytes and $5$ hops & Achieve sub-\SI{1}{\milli\second} latency without routing, coordination, and tight synchronization & Not compatible with low-power COTS radios, requiring custom HW/SW design\\
\hline
\end{tabularx}
\endgroup
\end{table*}

Although the protocols in Table~\ref{table5} show promising performance, they exhibit poor scalability behavior. Specifically, in these networks, the end-to-end network latency is very sensitive to the hop counts, which means that the end-to-end latency increases dramatically as the hop counts increase, thus limiting the deployment of large-scale IoT networks. For instance, in Glossy, each additional hop adds an extra \SI{500}{\micro\second} delay. The root cause of poor scalability and large latency increase per hop in the existing protocol is the store-and-forward switching paradigm, which requires the node to wait for the whole reception of the packet before decoding and retransmission. Consequently, each additional hop incurs a latency increase of at least the whole frame transmission time. In addition, the complicated coordination among relays and strict time synchronization are potentially required in the store-and-forward transmission, which is challenging for some energy-constrained IoT scenarios. To provide a quasi-deterministic latency regardless of hop counts and avoid the demanding coordination and synchronization, we propose to break the paradigm of store-and-forward switching and replace it with symbol-synchronous transmission in this article.

In our recent preliminary work, we performed early validation studies of RF-based symbol-synchronous communications~\cite{10590570, liu2024low}. Both studies explore the performance of symbol-synchronous transmissions using simulation. However, these simulations lack validation from experiments with actual RF transceivers, and the results are overly optimistic and fail to reflect practical constraints, such as the CFOs caused by imperfect clocks of transceivers. Motivated by this early work, we use SDR experiments in this work to calibrate the parameters in the simulation and explore the limitations of hardware in the design of symbol-synchronous protocols. Moreover, this article first specifically shows and analyzes how interference caused by symbol-level concurrent transmissions affects network reliability. At the same time, the interference is reproduced in simulation, thereby providing fairer evaluation.     

\section{RF-Zero-Wire design}
\label{sec3}
\subsection{System model}
\label{sec3_1}
We consider a scenario in which a source node broadcasts (floods) information to all other nodes in the network. For example, in self-organized cooperative robotic swarms, robots share information about their location, sensor data, and task statuses with all members of the swarm. To guarantee a high level of coordination, flooding should use a low-latency communication protocol. Since communication should happen over a fairly long range, relay techniques are utilized to forward frames to all nodes, especially those far away from the source node. Therefore, all other nodes not only take care of the detection of information symbols but also concurrently relay information to nearby nodes. In other words, each node serves as both a destination and a relay.

\begin{figure*}[!t]
\centering
\includegraphics[width=5.8in]{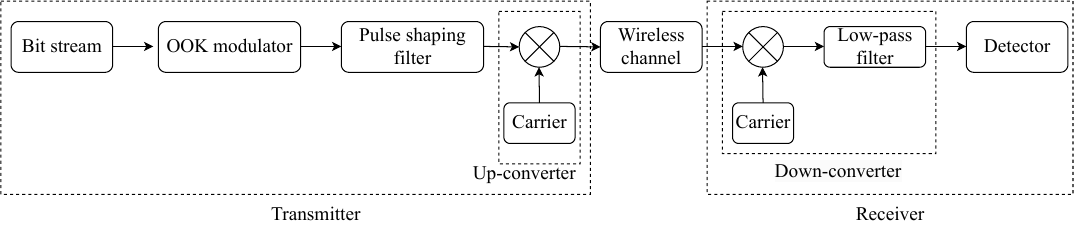}
\caption{Transceiver scheme}
\label{fig2}
\end{figure*}

Every node in the network is equipped with a transceiver capable of receiving, detecting, and relaying according to the symbol-synchronous transmission protocol proposed in this paper. The term \textit{symbol-synchronous} refers to the condition in which the nodes within the mutual communication range relay the same symbol during each symbol period. \mbox{Fig.~\ref{ss}} illustrates the general idea of symbol-synchronous transmission, where each relay node detects and relays the current symbol instantly after receiving a sufficient number of samples, rather than waiting until the entire frame is received, as is done in \mbox{store-and-forward} routing. Each transceiver consists of a pulse-based \mbox{on-off} keying (OOK) modulator, a pulse-shaping filter, an \mbox{up-converter}, a \mbox{down-converter}, and a customized \mbox{window-based} symbol detector. The \mbox{pulse-shaping} filter is used to limit the signal bandwidth. The transceiver design is depicted schematically in \mbox{Fig.~\ref{fig2}}, and further discussed in the remainder of this section.

\subsection{Pulse-based OOK modulation}
\label{sec3_2}
\begin{figure}[!t]
\centering
\includegraphics[width=3in]{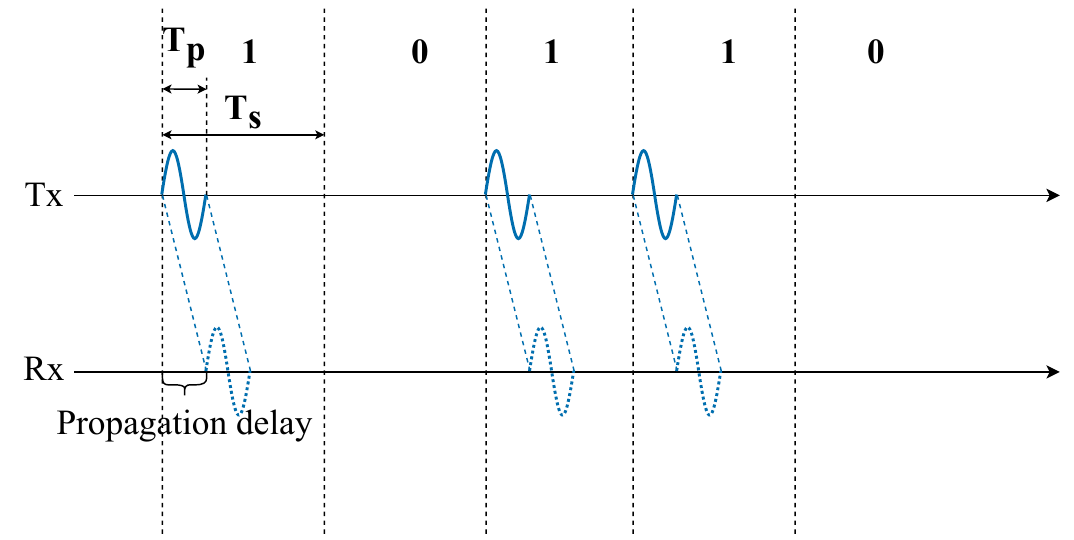}
\caption{Pulse-based OOK modulation}
\label{fig3}
\end{figure}

We apply the \mbox{pulse-based} OOK modulation scheme on the transmitter side. OOK modulation is a widely used technique in WSNs, especially in energy-constrained devices, like zero-energy devices~\cite{10615390}, and wake-up radios~\cite{pereira2020challenges}, due to its simple hardware implementation and low-complexity demodulation~\cite{10946880}. Moreover, OOK modulation can more effectively demodulate non-synchronized, overlapping pulses from concurrent transmissions, compared with other modulation schemes such as phase shift keying (PSK), where phase information is severely distorted after superimpositions of multiple signals from different paths and with various CFOs. To ensure that subsequent symbol transmissions do not interfere with each other, the pulse duration $T_p$ is kept shorter than the symbol period $T_s$. Each pulse is therefore followed by a silent ``guard'' period. When sending symbol $1$, the transceiver transmits a short pulse for duration $T_p \ll T_s$, as shown in \mbox{Fig.~\ref{fig3}}. When symbol 0 is sent, the transceiver remains silent during the entire symbol period $T_s$. A raised cosine filter is applied to shape the pulse to limit the bandwidth of the baseband signal and mitigate spectrum leakage. 

\subsection{Symbol-synchronous protocol}
\label{sec3_3}
This section introduces the proposed symbol-synchronous transmission protocol, designed to provide low \mbox{end-to-end} latency in wireless multi-hop networks. The core idea of the protocol is that it allows each node to relay the information symbol by symbol. More specifically, relay nodes forward each symbol immediately upon detection, without any coordination with neighboring nodes or buffering the message content. This protocol allows concurrent relaying without the need for explicit routing or flooding.

\begin{figure}[!t]
\centering
\includegraphics[width=2.5in]{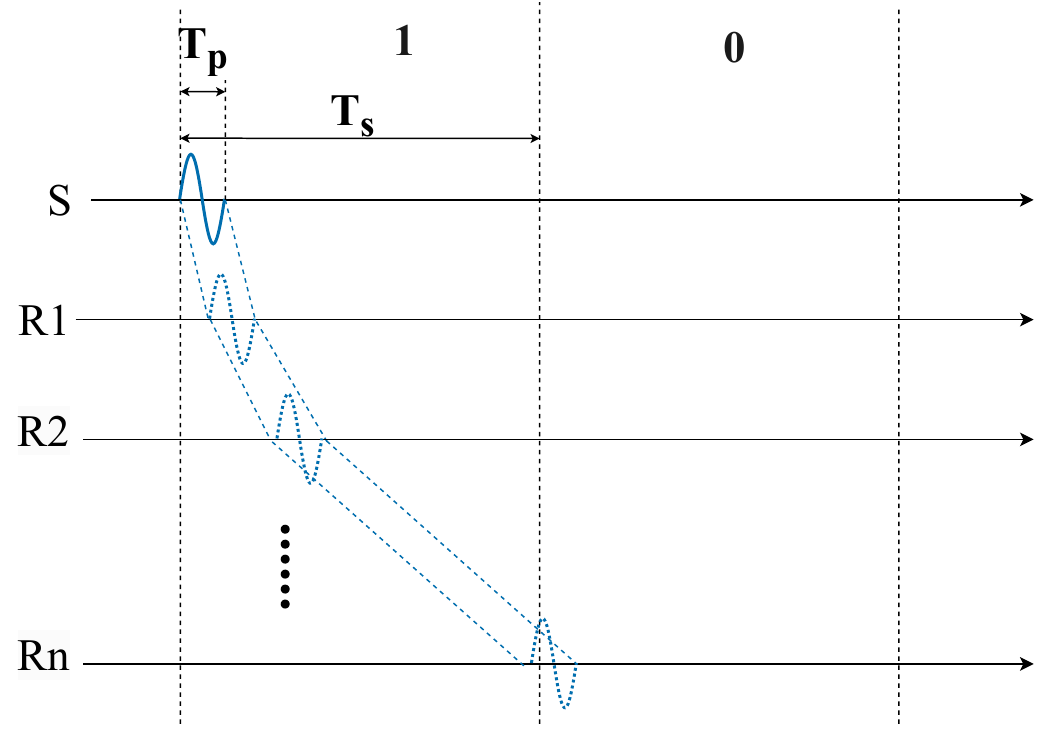}
\caption{Inter-symbol interference: symbol $1$ relayed by the node $R_n$ interferes with symbol $0$ sent by the source node $S$}
\label{fig5}
\end{figure}

Broadcast transmission unavoidably brings about some challenges. First, each relay node is likely to receive the relayed versions of the same symbol multiple times, which can be solved using a properly designed detector, where a sliding window is applied to trigger and terminate detection. Second, the relay nodes need to deal with \mbox{inter-symbol} interference (ISI). Since there is no predefined routing, they cannot distinguish the received signals from the current symbol or the previous symbol. Specifically, it is apparent that the nodes close to the source node, referred to as nearby nodes, detect the symbol earlier than those far away from the source node, referred to as distant nodes. Over time, when the nearby nodes start to receive and detect the next symbol, the distant nodes might still be relaying the previous symbol. In that case, the nearby nodes suffer from interference caused by the echo from the distant nodes, as shown in \mbox{Fig.~\ref{fig5}}.

Whether or not ISI can occur depends on the duration of $T_p$ and $T_s$. A large value of $T_s$ (relative to $T_p$) will ensure that all nodes in the network have forwarded a symbol before the next symbol is transmitted by the source. However, a smaller value of $T_s$ increases the data rate. As such, the configuration of $T_s$ entails a trade-off between reliability and data rate. If the number of hops $h_{\text{sym}}$ that can be covered within a symbol period $T_s$ is larger than the diameter of the network $h_{\text{max}}$ (i.e., the maximum hop count of the shortest path between any pair of nodes), ISI can be completely avoided. In our design, $T_p$ is kept fixed, while the value of $T_s$ can be varied to find the optimal trade-off between reliability and data rate.

\begin{figure}[!t]
\centering
\includegraphics[width=3in]{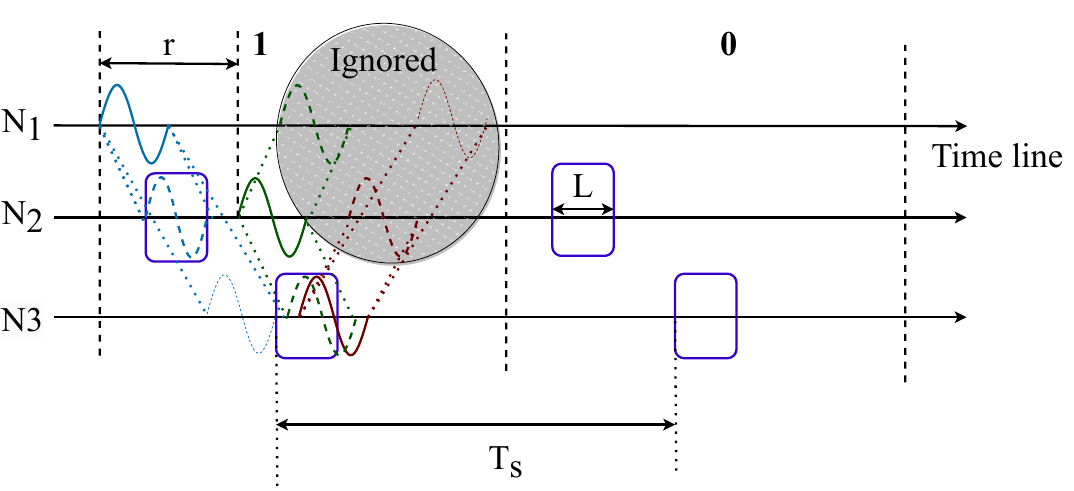}
\caption{Window-based symbol detection}
\label{fig4}
\end{figure}

Fig.~\ref{fig4} shows a simple example of the proposed relaying method. Here, $N_1$, $N_2$, and $N_3$ are three relay nodes that are mutually reachable. Firstly, $N_1$ successfully decodes symbol $1$, and instantly relays the symbol as a short pulse. After a time offset, which depends on the propagation delay, $N_2$ and $N_3$ receive the relayed symbol from $N_1$. Once they detect the symbol, they both relay it as well. Both $N_1$ and $N_2$ will receive echoes of the symbol they previously relayed. With proper configuration of $T_s$, it can be ensured that these echoes arrive within the same symbol period, allowing them to be easily ignored. As a 0 symbol is encoded as silence, no relaying happens when it is transmitted.

In terms of the network \mbox{end-to-end} latency offered by the \mbox{symbol-synchronous} protocol, it is largely determined by symbol period $T_s$. While the number of hops also has an effect, its impact is minimal, as each hop only increases latency by the propagation delay plus symbol detection time. This is the key advantage of the symbol-synchronous transmission compared to \mbox{store-and-forward} routing, where the latency increases by the sum of frame transmission time and waiting time for channel access per hop, resulting in a sharp increase of delay with the hop count. Let us define the maximum relay delay per hop $r \ll T_{s}$ as the sum of the time of signal propagation, hardware processing, and detection. 
Thus, the end-to-end latency $D$ of the network is constrained by 
\begin{equation}
    (n-1)T_s \leq D \leq (n-1)T_s + r h_{\max},
\label{eq2}
\end{equation}
where $n$ is the total number of bits in the frame. 

According to Equation~\eqref{eq2}, with this transmission pattern, each additional hop introduces only a delay of $r$, which is much smaller than $T_s$. In other words, the end-to-end latency of our network is bounded by the frame-transmission time, which means our network can provide a quasi-deterministic end-to-end latency, regardless of the number of hops. In contrast, for store-and-forward switching, the end-to-end latency $D_{\text{sf}}$ of the network is formulated by
\begin{equation}
    D_{sf} = h_{\max}\,(nT_s + T_{acc}),
\label{eq4}
\end{equation}
where $T_{acc}$ represents the waiting time for accessing the channel. 

Equation~\eqref{eq4} reveals that each additional hop in the store-and-forward routing contributes to a full frame transmission delay, as well as extra waiting time caused by MAC-layer processing and channel access procedure. 

\subsection{Detector design}
\label{sec3_4}
As mentioned previously, the symbol-synchronous protocol needs a tailored detector to accurately detect the symbol from the received signal, mitigating the interference of redundant copies of the same symbol (cf., Fig.~\ref{fig4}). Therefore, we design a window-based detector, where observation is limited to a short time slot called the detection window, whose duration is defined by the window length \mbox{$L$ ($T_p < L <T_s$)}. The transceiver only listens to the signals and makes decisions during the detection window. Once a symbol ($1$ or $0$) is successfully detected, the transceiver switches from reception to transmission and relays this symbol to neighboring nodes. The detection window will slide to the next position, exactly one symbol period $T_s$ further.

As the arrival time of the same symbol at different relay nodes varies with their distance to the source node, the positions of the detection window also differ among the relay nodes. Thus, a single $1$-bit is prepended to each frame as a preamble. If a fixed frame payload size is assumed, then a $1$-bit preamble is sufficient for nodes to detect the start of a new transmission, and synchronize their detection windows to the symbol start. In other words, the start of the symbol period in every relay is synchronized with the start of its detection window, rather than with that of the source node, thereby avoiding complicated explicit time synchronization. In order to detect the preamble, each relay node is assumed to be in continuous listening mode while it is not relaying a frame. Reducing the need for continuous listening, as well as a more robust preamble design, are both left for future work. Assuming the detection moment of the preamble bit at node $n$ is $t_n$, then the starting time of the detection window $t_{w,n}$ at $n$ is given by
\begin{equation}
    t_{w,n} = t_n - r - \tau,
\label{eq3}
\end{equation}
where $\tau$ is a potentially minor time offset introduced by the internal clock drift or other unpredictable factors. 
It is assumed that $L$ and $T_s$ are fixed and known by all nodes. Then, with the knowledge of $L$, $T_s$, and the calculated window position, each node can demodulate the frame symbol-by-symbol, without the need for a priori synchronization.

\begin{figure*}[!t]
\centering
\includegraphics[width=7in]{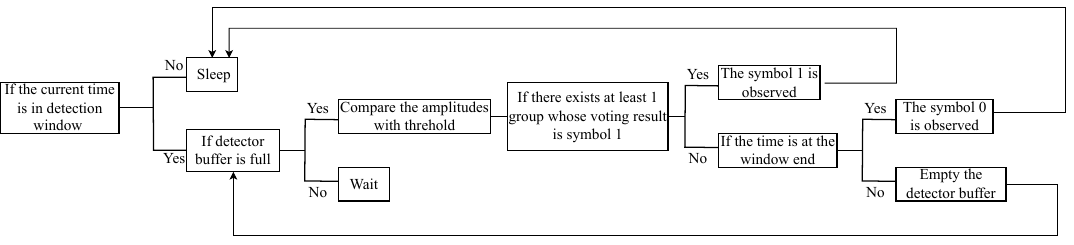}
\caption{Detection process block diagram}
\label{fig7}
\end{figure*}

Within the detection window, a \mbox{voting-based} detection method is applied multiple times to improve the detection accuracy. The entire process is shown in Fig.~\ref{fig7}. First, we select a fixed buffer size at the transceiver to store the received I/Q samples. When the buffer is full, the stored samples are divided into groups of k samples each. Each group is then sequentially sent to a comparator, where the amplitudes of the samples are compared against a predefined threshold. In this work, the threshold is set to $9$~dB higher than the noise floor of the ambient noise collected over a fixed period prior to each transmission. For each group, a voting method is applied: if more than half of the sample amplitudes in the group exceed the threshold, the symbol~$1$ is considered to be detected. If none of the groups in the buffer are detected as symbol $1$, the buffer is cleared, and new IQ samples fill the buffer for the next detection. If the detector fails to detect symbol $1$ at the end of the window time, we determine that symbol 0 is detected.

The length of the detection window determines the detection latency and power consumption on one hand, and the trade-off between the symbol error rate~(SER) of symbols $1$ and $0$ on the other hand. A longer detection window enables more detection attempts, but also increases the radio-on time (and thus power consumption). In terms of robustness, more detection attempts improve the detection probability of 1-symbols by reducing the probability of the CFO-induced beating effect causing destructive interference during the entire detection window. On the other hand, a longer detection window increases the chance of noise being wrongfully detected as a $1$-symbol. It should be noted that the window detector will not affect the end-to-end network latency significantly, since the end-to-end latency is bound by the frame transmission time, which depends on the symbol duration instead of the detection window duration, as shown in Equation~\eqref{eq2}.

\section{Hardware calibration}
\label{sec4}

To develop a realistic- and hardware-aware simulation implementation, we calibrate the key parameters such as transmit power, maximum transmission distance, carrier frequency, and bandwidth experimentally using a simplified implementation on a \mbox{software-defined radio (SDR)} platform. After suitable values for these parameters are selected through hardware experiments, they are applied to our MATLAB-based simulation framework (cf., Section~\ref{sec5}). Furthermore, in symbol-synchronous transmission, several relay nodes may transmit the same symbol at overlapping times, which we call concurrent transmission. Such concurrent transmissions result in overlapping radio signals at the receiver. The overlapping signals can potentially increase the signal strength, thus contributing to more accurate detection at the receiver. This is referred to as constructive interference. However, signal overlap can also result in destructive interference, reducing the signal strength. This contrasts prior attempts at symbol-synchronous communications that rely on optical wireless communication, where interference is always constructive~\cite{10.1145/3384419.3430897}. The potential occurrence of destructive interference during concurrent symbol transmissions is also assessed based on the hardware experiments. 

\subsection{SDR platform}
\label{sec4_1}
For calibration, we select the ADALM-PLUTO SDR (Rev. C) to validate the transceiver logic~\cite{ADALM_PLUTO_Datasheet}. Compared with other mainstream SDRs, such as the National Instruments USRP series, it offers a low-cost solution for prototyping and experimentation, in line with the capabilities of low-cost IoT devices. It integrates a radio front-end based on the AD9363 chip~\cite{AD9363_Datasheet} and a Zynq-7010 \mbox{system on chip~(SoC)}~\cite{Zynq7000_DS190}, combining an ARM-based processor and compact Field-Programmable Gate Array~(FPGA). The ARM-based processor of the Zynq-7010 SoC enables the module to run a customized standalone application without connecting to a host computer for control and processing. In addition, the Zynq-7010 SoC performs both \mbox{high-level} software control and low-level digital signal processing. The AD9363 chip embedded on the ADALM-PLUTO SDR offers a wide carrier frequency range from \SI{325}{\mega\hertz} to \SI{3.8}{\giga\hertz} and supports the tunable channel bandwidth up to \SI{20}{\mega\hertz}~\cite{AD9363_Datasheet}. In addition, a W1059 antenna~\cite{W1059_Datasheet} is connected to the SDR, in order to transmit and receive the analog signals. A host computer with the Ubuntu 22.04.4 LTS operating system is used to store and process the received signals.

\subsection{Experimental setup}
\label{sec4_2}
Owing to the Zynq-7010 SoC providing an on-board Linux operating system, we can easily interact with the SDRs via SSH. Accordingly, we respectively developed C programs for the transmitter and receiver to implement the functions of signal transmission and reception. These programs were cross-compiled into executable files and deployed to the SDRs, enabling them to automatically transmit and receive signals upon power-up. At the same time, for each transmitter, a predefined binary data file containing complex baseband IQ samples mapping the transmitted symbols is constructed and stored in the SDRs.

Two sets of experiments with the Pluto SDRs in a spacious indoor laboratory with minimal obstructions are conducted. The room is not shielded, and background interference may thus occur, in line with real-world deployment. The reliability of reception is evaluated and compared under these two experimental setups. In the first setup, one SDR transmits signals while another SDR receives them at different positions. In the second setup, three SDRs located at different positions simultaneously transmit identical signals, which are received as overlapping signals by a fourth SDR.
\begin{figure}[!t]
\centering
\includegraphics[width=3in]{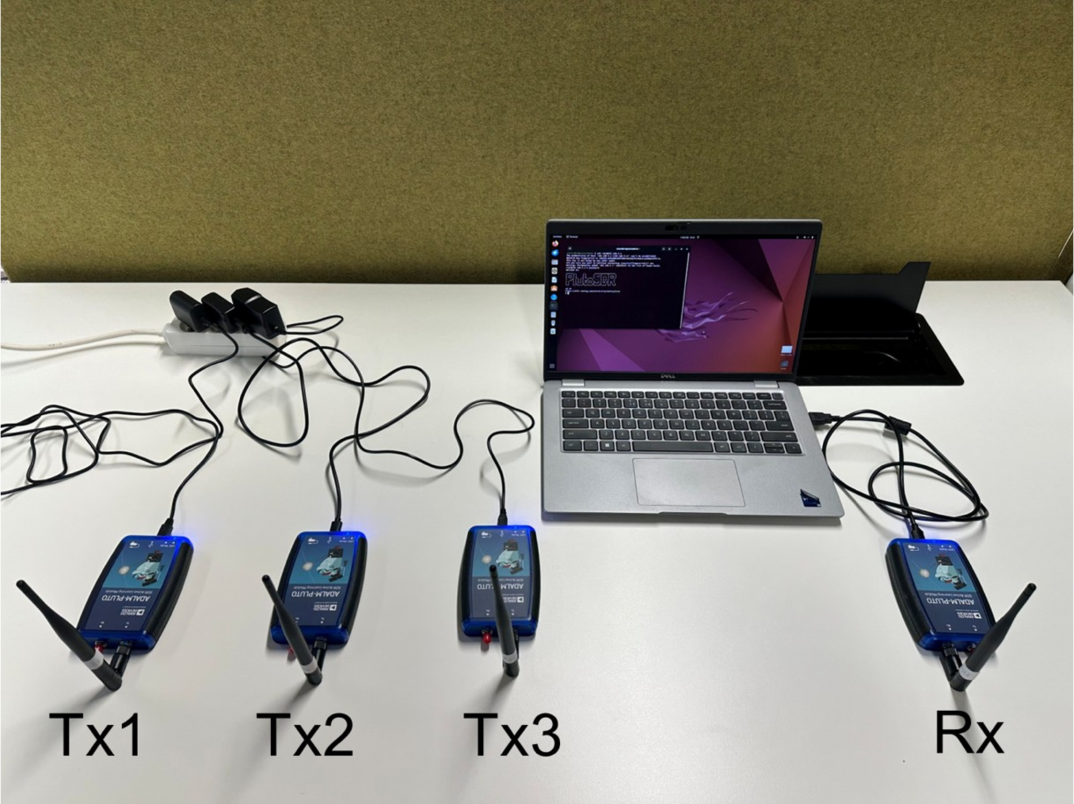}
\caption{Experimental setup with ADALM-PLUTO SDRs}
\label{hw_setup}
\end{figure}
Here, the three transmitters, denoted by TX1, TX2, and TX3, are connected to a shared power supply, but are not time-synchronized. The receiver, denoted by RX, is connected to the host computer, as shown in Fig.~\ref{hw_setup}. At the beginning, the power supply is turned on and triggers the three transmitters simultaneously. The receiver is enabled by the host and collects the I/Q samples of the combined signal from all three transmitters. Finally, all the collected data is loaded onto the computer and post-processed using the detection algorithm proposed in Section~\ref{sec3_4}.

\begin{figure}[!t]
\centering
\includegraphics[width=3.4in]{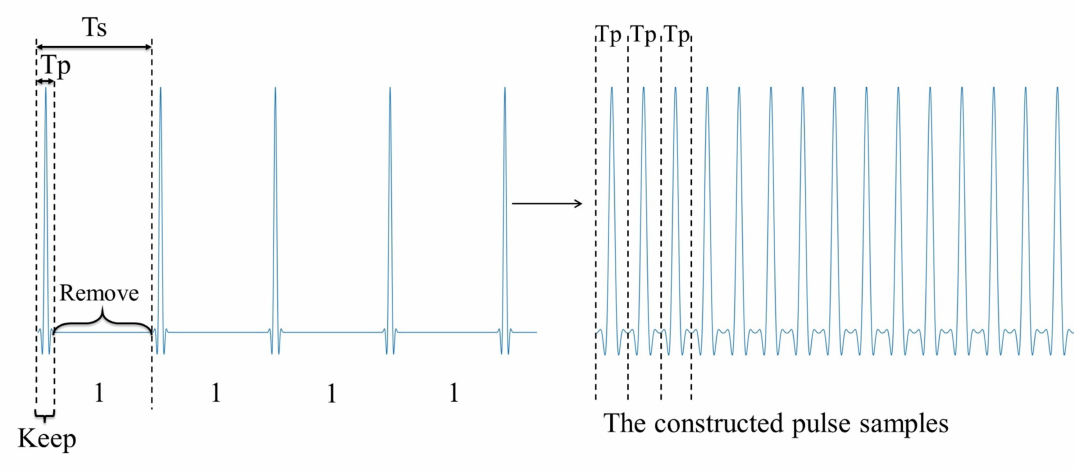}
\caption{The construction of the pulse signal used in the hardware calibration experiments}
\label{fig9}
\end{figure}

In these experiments, we are solely interested in the ability to successfully decode the pulses that are generated for each $1$-symbol. As such, the transmitters are configured to transmit a sequence of pulses of duration $T_p$ one after the other, while the guard period of length $T_s - T_p$ is omitted (cf., Fig.~\ref{fig9}). This allows us to determine successful reception probability as a function of distance and transmit power, as well as study the effects of constructive and destructive interference of overlapping pulses. As the transmitters are not time synchronized, a wide range of randomly overlapping pulses, with varying phase shifts, will occur throughout the experiments. This is in line with the proposed protocol, where the relays are also not synchronized. At the start of each experiment, all transmitters remain off and the receiver collects ambient noise for $1$~second to determine the noise floor and properly configure the amplitude threshold for symbol detection.

Several other parameters were configured as follows, and the same configuration is applied to the subsequent simulations (cf., Section~\ref{sec5}). The transmit power was set to $0$~dBm, which is commonly used in low-power IoT networks~\cite{yang20190}. 
For the experiments, the frequency range \mbox{2483.5-\SI{2500}{\mega\hertz}} was used, as it did not suffer from interference from Wi-Fi or other technologies using the ISM band. Local regulation for this band specifies that only a bandwidth of less than \SI{3}{\mega\hertz} can be used for a short time and low power. As such, we set the bandwidth to \SI{2.8}{\mega\hertz}, which is consistent with well-established wireless standards in multi-hop sensor network applications, such as IEEE 802.15.4~\cite{natarajan2016analysis}. Accordingly, the pulse duration $T_p$ was set to \SI{3}{\micro\second} and the sampling rate of the transmitter to \SI{20}{\mega\hertz}.
At the receiver side, the output gain was fixed. As the hardware experiments do not involve multi-hop transmissions nor include a silent guard interval, the window length and buffer size of the detector are both configured to be equal to the pulse duration. The number of samples $k$ per group for voting is fixed to $10$, corresponding to the number of samples included in a pulse before the pulse shaping filter. All essential parameters are listed in Table~\ref{table1}. 

\begin{table}[!t]
\caption{Parameters used in hardware experiments\label{table1}}
\centering
\begin{tabular}{c|c}
\hline
\textbf{Parameter} & \textbf{Value}\\
\hline
Center carrier frequency & \SI{2491}{\mega\hertz}\\
Sampling rate & \SI{20}{\mega\hertz}\\
Transmit power & $0\,\mathrm{dBm}$\\
Antenna gain~\cite{W1059_Datasheet} & $5\,\mathrm{dBi} \pm 1\,\mathrm{dB}$\\
Gain control mode of the Pluto receiver & Manual\\
Manual gain of the Pluto receiver & \SI{60}{\dB}\\
Pulse duration & \SI{3}{\micro\second}\\
Bandwidth & \SI{2.8}{\mega\hertz}\\
k & $10$~samples\\
\hline
\end{tabular}
\end{table}

\begin{figure}[!t]
\centering
\includegraphics[width=3.4in]{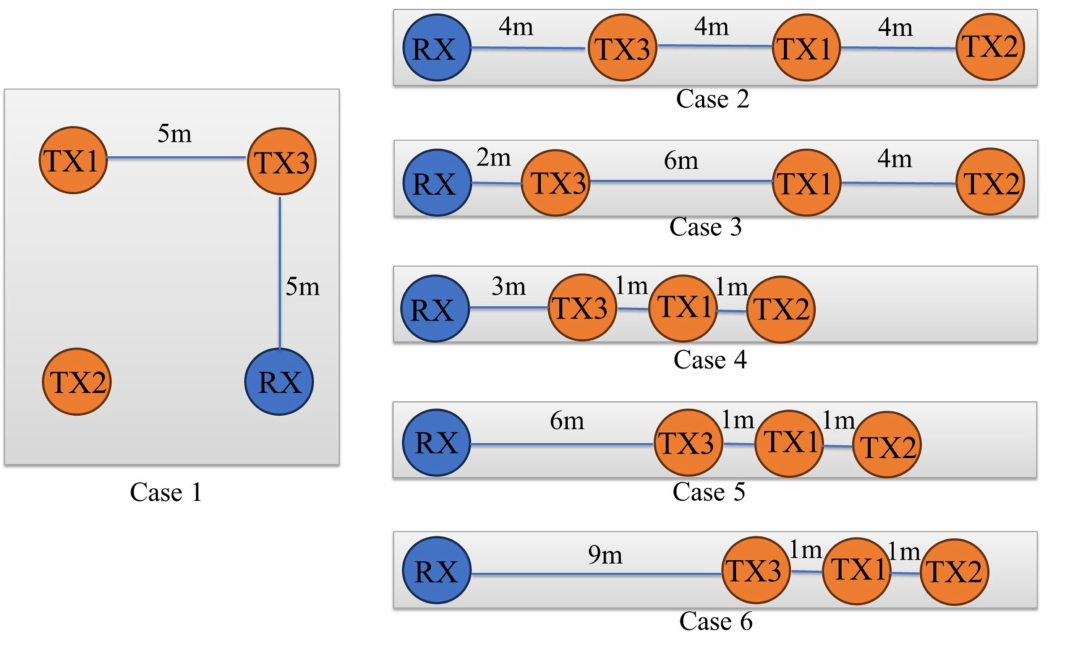}
\caption{Network topologies used in the hardware calibration experiments}
\label{fig10}
\end{figure}

Network performance was evaluated on $6$ different network topologies, as shown in Fig.~\ref{fig10}. Every topology consists of the $4$ nodes placed in the large indoor laboratory with a size of \SI{72}{\square\meter}. As was mentioned earlier, on each network topology, the experiment comprises two phases: communication between a single transmitter and a single receiver, and communication between three transmitters and a single receiver. In the first phase, each transmitter is activated individually.  Specifically, we first power on TX1 and keep TX2 and TX3 off, then activate the receiver to collect I/Q samples. After completion, TX1 is turned off. This procedure is then repeated for TX2 and TX3, respectively. In the second phase, all three transmitters are turned on simultaneously, and the receiver collects the overlapping signals. The experiment is repeated five times, and in each trial, the receiver collects \SI{1}{\second} of data (around 333333 pulses). Finally, the detection algorithm (cf., Section~\ref{sec3_4}) is applied on the collected data, and the SER of the received signals, indicating the probability of missing a $1$ symbol due to destructive interference, ambient noise, or interference from other sources, is calculated. 

\subsection{Experimental results and analysis}
\label{sec4_3}

\begin{table*}[!t]
\sisetup{
  scientific-notation = true,
  exponent-product = \times,
}
\centering
\caption{Network Reliability Evaluation For Different Network Topologies and Transmission Schemes on transmitted \num{1.67e6} bits}
\label{table2}
\begin{tabular}{l|ccc|c}
\toprule
\multirow{2}{*}{\textbf{Topology}} & \multicolumn{3}{c|}{\textbf{SER during Phase $1$}} & \multicolumn{1}{c}{\textbf{SER during Phase $2$}} \\
                  & \textbf{TX1} & \textbf{TX2} & \textbf{TX3} & \textbf{TX1 + TX2 + TX3} \\
\midrule
Case $1$       & \num{0} & \num{0} & \num{0} & \num{9.05e-6}\\
Case $2$       & \num{1.80e-6} & \num{4.62e-1} & \num{0} & \num{0}\\
Case $3$       & \num{1.25e-6} & \num{3.98e-1} & \num{0} & \num{0}\\
Case $4$       & \num{0} & \num{0} & \num{0} & \num{0}\\
Case $5$       & \num{0} & \num{0} & \num{0} & \num{4.44e-6}\\
Case $6$       & \num{1.92e-4} & \num{1.61e-1} & \num{2.18e-5} & \num{1.31e-4}\\
\bottomrule
\end{tabular}
\end{table*}
For each network topology and transmitter configuration, the SER is displayed in Table~\ref{table2}. For the transmission phase $1$, where there is only one transmitter and one receiver, the transmission reliability mainly depends on the distance between them. When the transmission distance is longer than \SI{8}{\meter}, detection errors start occurring, as evidenced by the increased SER when using TX1 in cases $2$ and $3$. More severely, the reception performance is further degraded for TX2 in case $6$, where a SER of \SI{16.1}{\percent} is observed due to the longer transmission distance of \SI{11}{\meter}. A transmission distance of \SI{12}{\meter} further increases the SER to \SI{46.2}{\percent}, which would be considered too high for nearly any application. Therefore, a maximum distance of \SI{11}{\meter} between any node and its closest neighbour will be ensured in the subsequent simulations.

Comparing the SER during phase $1$ and phase $2$ shows that multiple concurrent transmissions will cause additional symbol loss. Specifically, in the phase $1$ experiment of case $1$, no detection errors occur due to the near distance of \SI{5}{\meter}. However, if the three transmitters are enabled simultaneously, some detection errors occur. The same phenomenon can be observed in case $5$. These results show that destructive interference occurs during concurrent transmissions. 

\begin{figure}[!t]
\centering
\subfloat[Single transmitter TX2 (Phase 1)]{\includegraphics[width=3.4in]{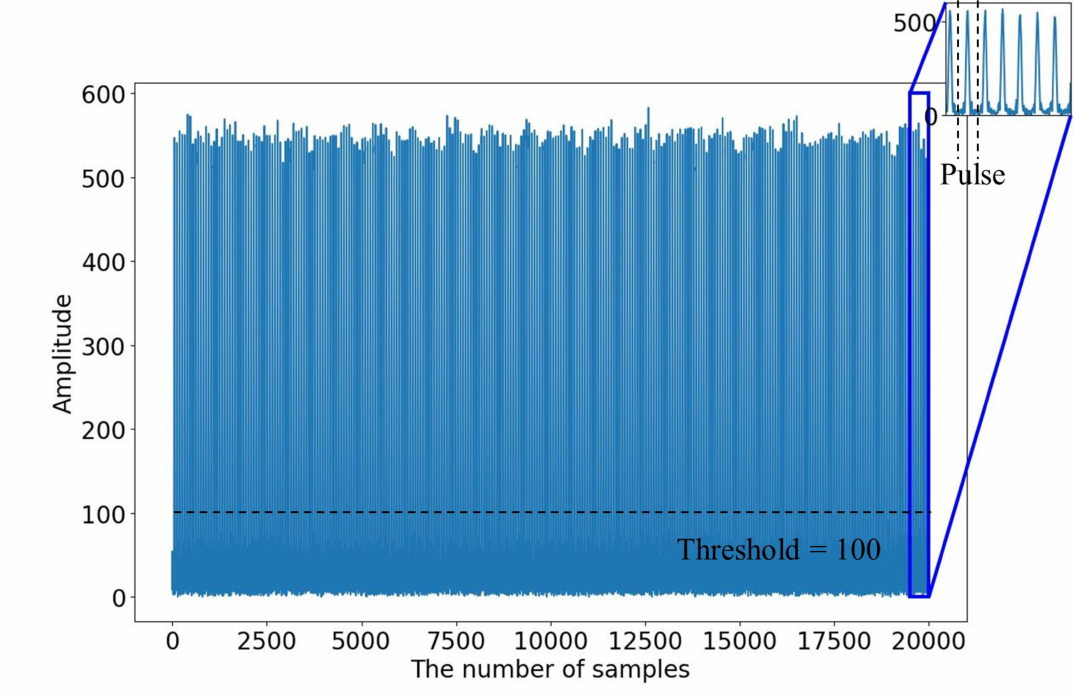}}
\label{fig11_2}
\hfil
\subfloat[Three simultaneous transmitters TX1, TX2, and TX3 (Phase 2)]{\includegraphics[width=3.2in]{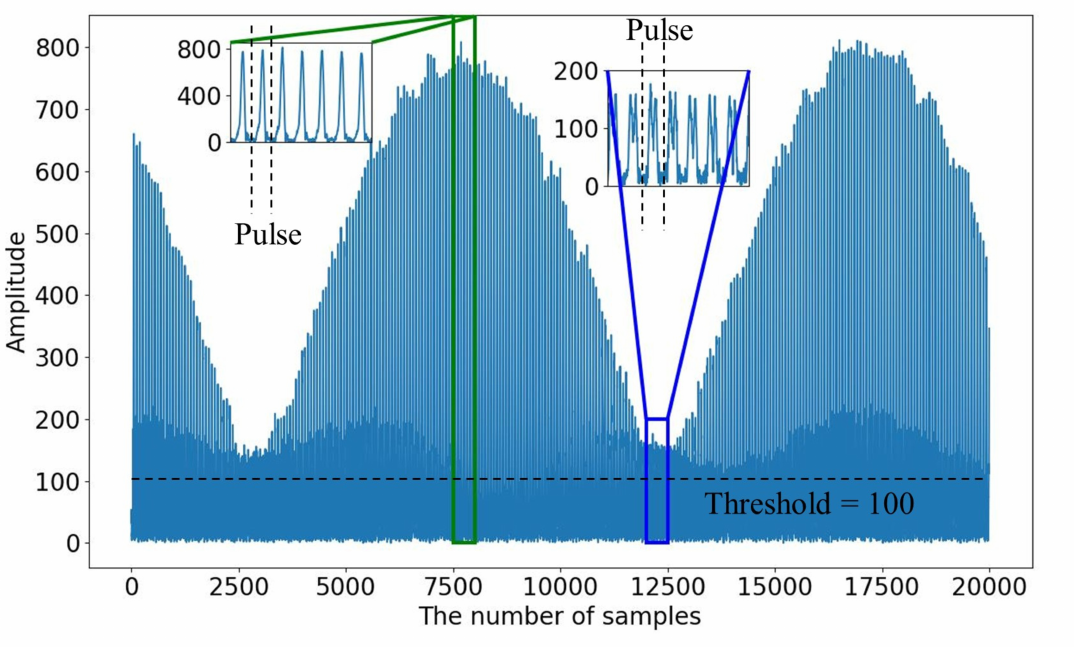}}
\label{fig11_1}
\caption{The received signals on the Case $1$ network topology}
\label{fig11}
\end{figure}

To gain deeper insights into the interference caused by concurrent transmissions, we visualize the received signals. Fig.~\ref{fig11} shows a segment of the received signal samples in the experiment phase $1$ and phase $2$ on the first network topology (case $1$), showing the impact of concurrent transmissions. Note that the amplitudes in Fig.~\ref{fig11} are the digitized value output by the analog-to-digital converter~(ADC) based on the received signal strength, rather than the signal’s true physical amplitude. In the transmission with a single transmitter, the strength of the received signal remains nearly constant with slight fluctuations caused by ambient noise, under the same transmission distance and environment. In contrast, the transmission involving three transmitters results in alternating increases and decreases in signal strength due to constructive and destructive interference. This phenomenon is also known as the beating effect~\cite{10.1145/3604430}. Such fluctuation exhibits short-term periodicity. This periodic fluctuation is mainly caused by CFOs among multiple transmitters. Although identical carrier frequencies were initially configured, carrier frequency drift occurs over time as a result of internal clock instability, which is affected by multiple factors, such as temperature variation and the aging of hardware components~\cite{10.1145/3604430}. The fluctuation period, also known as the beating period, depends on the difference in CFOs, which change over time. That is why the fluctuation is not strictly periodic in the long term. The destructive effects of the CFO among overlapping transmitters will be studied further in the system-level simulations in the next section.

\section{System-level simulation}
\label{sec5}

Based on the observations and lessons learnt from the hardware calibration experiments, a simulation framework was created using MATLAB. This framework offers greater flexibility and scalability, supporting in-depth analysis of the effects of various system and environmental parameters, as well as evaluation at a larger scale.

\subsection{Simulation setup}
\label{sec5_1}
\begin{figure}[!t]
\centering
\includegraphics[width=2.5in]{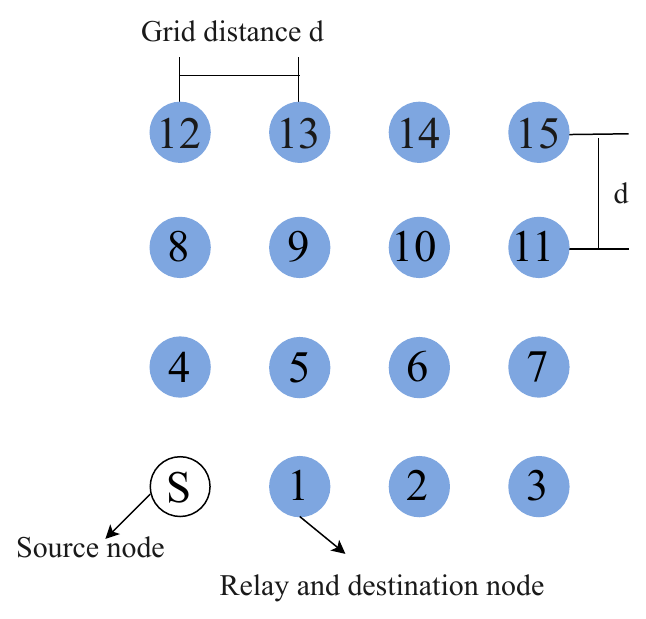}
\caption{The network topology used in the MATLAB simulation}
\label{fig12}
\end{figure}

In the simulation, we consider a multi-hop network where nodes are statically located in a regular lattice with a fixed horizontal and vertical grid distance of $d$, as illustrated in Fig.~\ref{fig12}.
There is one source node in the bottom-right corner, and all other nodes have the roles of both relay and destination nodes. 
The TGac channel model~\cite{ieee2009tgac}, proposed originally for Wi-Fi networks, is used to simulate the signal transmission environment, where the multipath fading characteristics, large-scale fading effects including pathloss and shadowing, and Doppler shift are simulated through clustered delay profiles and spatial correlation. Furthermore, the TGac model pre-defines six representative channel profiles (Models A–F) to simulate different propagation environments, ranging from small indoor offices to large open spaces and outdoor scenarios. In this article, TGac model D is selected as the default channel model for the simulations unless specified otherwise, as it is tailored for large indoor open spaces with line-of-sight~(LOS) conditions, matching our hardware-based experimental setup. 

To maintain consistency with the hardware experiments, all the parameters in Table~\ref{table1} are identically applied in the simulation. For our window-based detector, the parameters of the receiver buffer size and window length are different from those used in the hardware experiments, as the transmitted signal shapes are different between simulations and hardware experiments~(i.e., the transmitted pulses in hardware experiments are without a silent guard period). We set the receiver buffer size to $100$ samples and the window length to $200$ samples, which corresponds to \SI{10}{\micro\second}. With the aim of low latency, the buffer size should be as small as possible to decrease the relay time $r$. We set the buffer size to $100$ samples, which is slightly larger than the pulse size of $60$ samples, to account for time drift and multipath reflections. We assume that the hardware processing time is sufficiently small to be neglected in the simulation. Then the minimum relay time $r$ corresponds to the time needed to fill the buffer, i.e. 

\[ r = \frac{\text{buffer size}}{\text{sampling rate}} = \frac{100}{2 \times 10^{7}} = \SI{5}{\micro\second}.\] 

In addition, when selecting the window length, there is a \mbox{trade-off} between the accuracy of detection of $0$-symbols and $1$-symbols. Specifically, a longer window contributes to the greater likelihood of mistakenly detecting symbol $0$ as symbol $1$, while increasing the detection accuracy of symbol $1$ and robustness to clock drift. The window length of $200$ samples, allowing for two detection attempts per symbol, ensures a relatively balanced detection accuracy and power consumption between symbol $1$ and symbol $0$. We set the noise power to $-60$~dBm to ensure that the maximum transmission distance in the simulation is aligned with that measured during hardware experiments. The clock drift $\tau$ is set to \mbox{\SI{0.5}{\micro\second}} according to the maximum offset that we observed in hardware experiments and fixed for every node. With the longer window, buffer size, and compensation for clock drift, the window-based detector can robustly capture the pulse, even in the presence of timing offsets in heterogeneous IoT environments. Other critical parameters, such as data rate, grid distance, and the number of nodes, vary in different experiments. They are specifically described in the following sections. All simulation-specific parameters are listed in Table~\ref{table3}.

The transceiver implementation generally follows the design highlighted in Fig.~\ref{fig2}. First, the transmitter generates a uniformly distributed random bit stream. Every symbol is represented by one bit of data. After that, each symbol is sent to the OOK modulator implemented as a custom function, followed by a \mbox{pulse-shaping} filter with the roll-off factor set to $0.5$~\cite{mukumoto2014realization}, implemented using a built-in MATLAB object. Then we multiply the carrier by the output of the \mbox{pulse-shaping} filter. Through the TGac wireless channel, the signal is sent to the receiver. At the receiver side, the carrier is multiplied again by the received signal, which goes through a low-pass filter. Finally, a customized function implementing the symbol detector is applied. It is assumed that all nodes in the network are equipped with the same transceiver. Also, all parameters listed in Table~\ref{table1} and Table~\ref{table3} remain consistent in all simulation experiments. 

\begin{table}[!t]
\caption{Parameters used in the MATLAB experiments\label{table3}}
\centering
\begin{tabular}{c|c}
\hline
\textbf{Parameter} & \textbf{Value(units)}\\
\hline
Noise power & $-60$~dBm\\
Window length & $200$~samples (\SI{10}{\micro\second})\\
Receiver buffer size & $100$~samples\\
Detection attempts per symbol & $2$\\
Network topology & Mesh\\
Default channel model & TGac model D\\
Maximum clock drift $\tau$ & \SI{0.5}{\micro\second}\\
Roll-off factor of pulse shaping filter & $0.5$\\
\hline
\end{tabular}
\end{table}

Network performance is evaluated considering $3$ evaluation metrics: latency, reliability, and scalability.

\subsection{Network latency evaluation}
We construct a multi-hop network involving $25$ nodes, where the grid distance $d$ is fixed at \SI{5}{\meter}. We transmit $1500$ frames of fixed size, considering values from the set \{$8$, $16$, $32$, $64$, $128$, $256$, $512$, $1024$\}~bits. Various data rates from the set \{$100.0$, $66.7$, $50.0$, $40.0$, $33.3$\}~kbps are tested. The average network \mbox{end-to-end} latency per frame is calculated. Here, the data rate represents the number of symbols sent by the source node per second, which is equal to the inverse of $T_s$. The \mbox{end-to-end} latency is defined as the duration from the initiation of transmission by the source node of the first symbol to the successful reception and detection of all symbols in the frame by all destination nodes.

\begin{figure}[!t]
\centering
\includegraphics[width=3.5in]{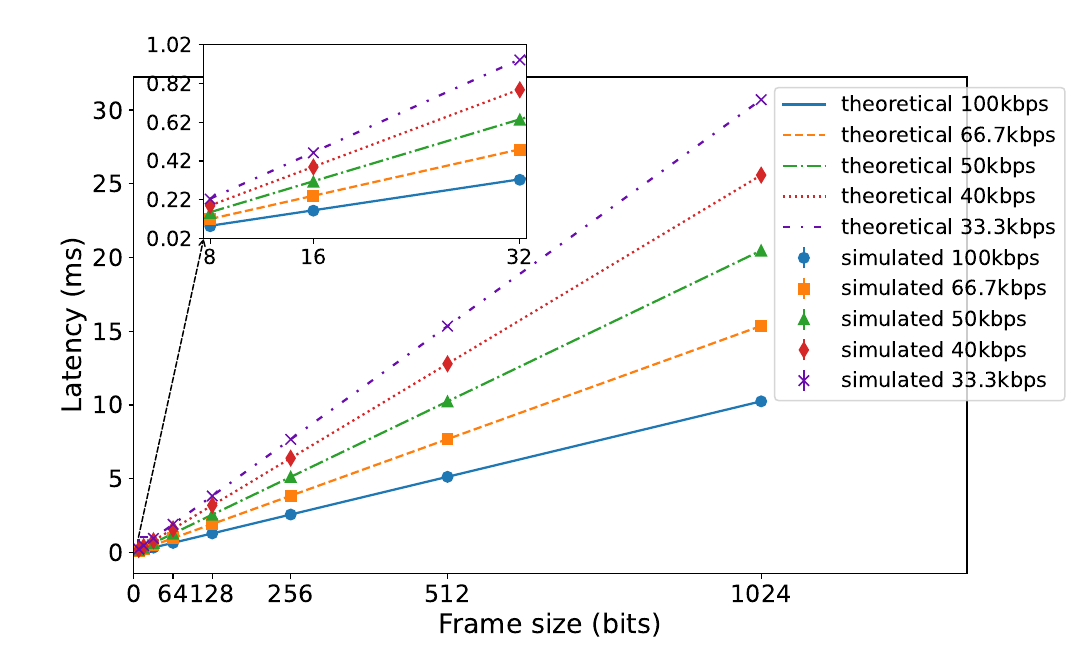}
\caption{Network end-to-end latency as a function of the frame size and data rate}
\label{fig13}
\end{figure}

As illustrated in Fig.~\ref{fig13}, the simulated end-to-end network latency complies well with the corresponding theoretical values formulated in Equation~\eqref{eq2}, and the network \mbox{end-to-end} latency is proportional to the data rate and the frame size. For instance, when the data rate is $100$~kbps, Fig.~\ref{fig13} shows that the average end-to-end latency for transmitting a $512$-bit frame is approximately \SI{5}{\milli\second}. The corresponding frame-transmission time is \SI{5.12}{\milli\second} ($\frac{512}{10^{5}} \times 10^{3} = \SI{5.12}{\milli\second}$), which is very close to the simulated end-to-end latency. This consistency further validates our theoretical conclusion: the end-to-end latency of our network is bounded by the frame-transmission time in multi-hop environments, which is also a key novelty of our proposed protocol. Under the transmission of a small frame, with a length below $32$~bits~($4$~bytes), the \mbox{end-to-end} latency is in the sub-millisecond order. For a relatively big frame of $1024$~bits ($128$~bytes), an \mbox{end-to-end} latency of $10$ to $30$ milliseconds can be achieved, depending on the data rate.

\begin{figure}[!t]
\centering
\includegraphics[width=3in]{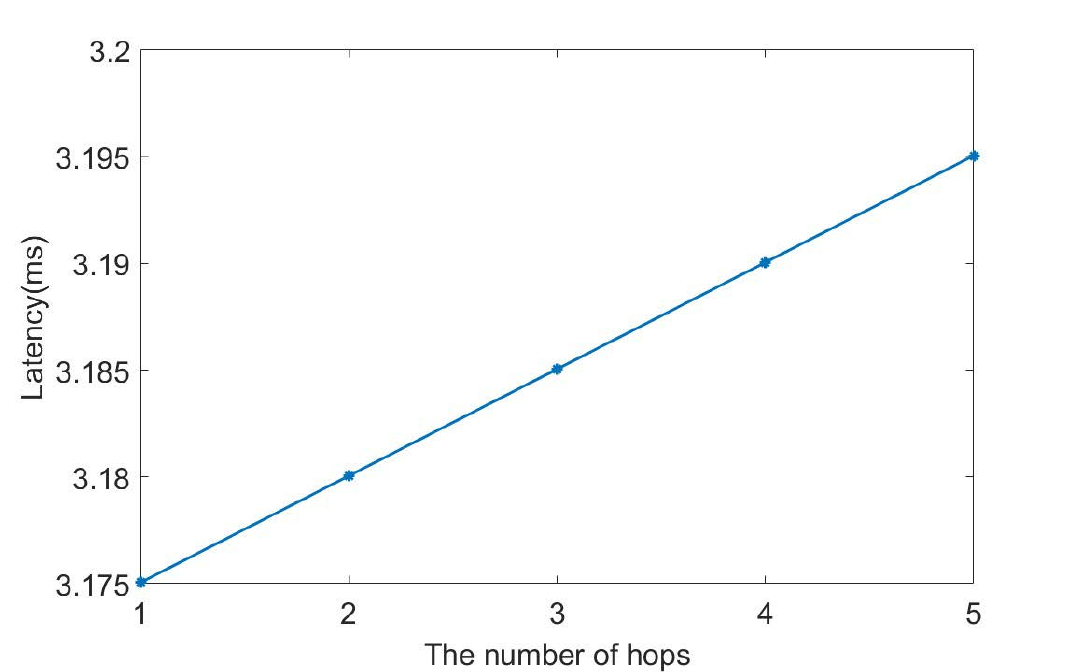}
\caption{Network end-to-end latency as a function of the number of hops}
\label{fig14}
\end{figure}

In addition to the \mbox{end-to-end} latency for different frame sizes, we also investigate how the latency scales with an increasing number of hops. We expand the network to $144$ nodes ($12$-by-$12$) to increase the hop count over the network. The grid distance $d$ of \SI{5}{\meter} remains the same, and the data rate is set to \mbox{\SI{40} kbps}. Fig.~\ref{fig14} shows the latency behavior as a function of the number of hops in the transmission of a $128$-bit frame. Latency increases linearly by \SI{5}{\micro\second} per extra hop, which is two orders of magnitude lower than the \SI{500}{\micro\second} per-hop delay increment in Glossy~\cite{5779066}, where store-and-forward forwarding is used. In addition, in this article, the delay increment per hop for the transmission of a $128$-bit frame is only about \SI{0.16}{\percent} of the overall end-to-end latency, which means the network end-to-end latency increases insignificantly as the hop count increases and exhibits near-deterministic and bounded behavior even over many hops. In contrast, the deterministic network end-to-end latency is challenging to achieve in conventional multi-hop wireless networks limited by the store-and-forward paradigm, where network end-to-end latency rapidly increases as the hop count increases. For instance, Glossy introduces around \SI{28}{\percent} delay increment per hop for the transmission of a $64$-bit frame. This increment is due to the extra relay time~$r$ introduced by each extra hop. These scaling properties pave the way for the deployment of large-scale distributed low-latency IoT networks, as required, for example, by cooperative robotic swarms.

\subsection{Network reliability evaluation}
To study the network communication reliability, we calculate the bit error rate (BER) based on a \mbox{$16$-node} network~($4$-by-$4$) with grid distance of \SI{5}{\meter} and data rate of $40$~kbps. Hardware experiments show that multiple concurrent transmissions with various CFOs result in alternating occurrences of destructive and constructive interference. Such interference lowers the detection accuracy. Therefore, we leverage the simulation platform to specifically explore the effect of CFOs on the symbol-synchronous transmission. Specifically, for the transceiver of every node, we generate a random CFO value within a certain range and add it to the carrier frequency of \SI{2491}{\mega\hertz}. Considering the hardware circuit instability, the CFOs are changed every second to mimic the hardware behavior in the simulation. We tested different CFO variation ranges from $\pm1\,\mathrm{Hz}$ to $\pm60\,\mathrm{KHz}$.
For every CFO variation range, we transmit data for $15$ seconds, and \mbox{$40$K} bits are transmitted and detected every second. After every one-second transmission interval, we calculate the BER of every node before the CFOs change. 

\begin{figure}[!t]
\centering
\includegraphics[width=3in]{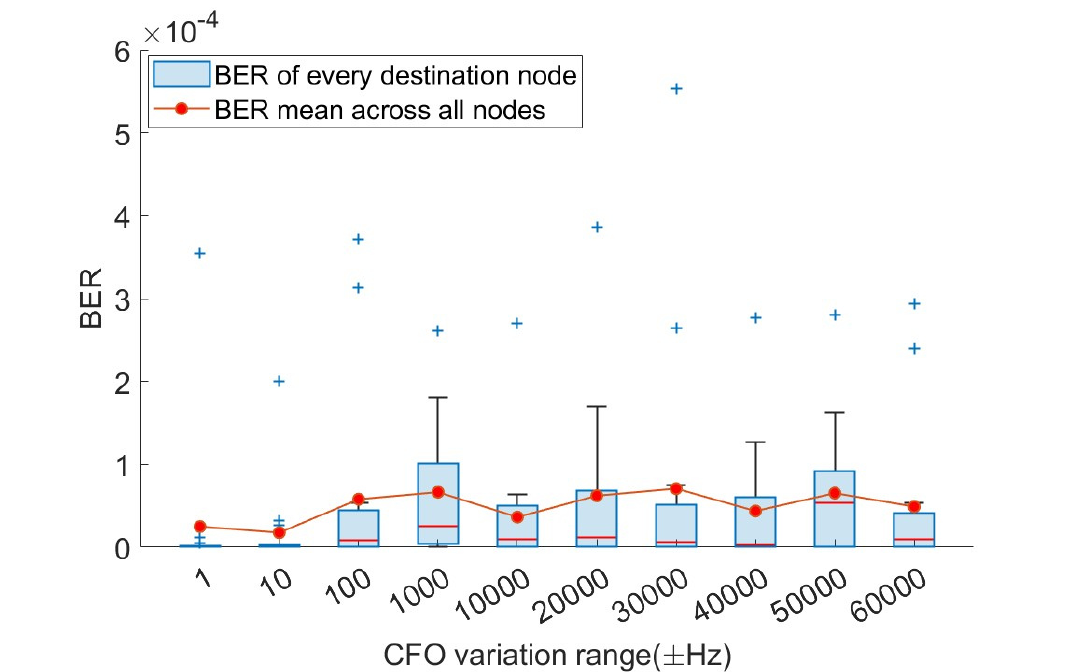}
\caption{Average bit error rate measured across all nodes as a function of CFO variation range}
\label{fig15}
\end{figure}

As for the different CFO variation ranges, the average BER of every destination node in the \mbox{$15$s-transmission} and the BER mean value across all destination nodes are shown in Fig.~\ref{fig15}. In \mbox{Fig.~\ref{fig15}}, the network BER, represented by the mean value of the BER across all nodes, fluctuates from \SI{0.002}{\percent} to \SI{0.007}{\percent}. When the BER of every destination node and the mean value of the BER across all nodes are considered jointly, it becomes evident that the magnitude of network-wide BER is not strongly correlated with the CFO variation range. Specifically, as illustrated in \mbox{Fig.~\ref{fig15}}, even under a small CFO variation range~(e.g., $\pm1\,\mathrm{Hz}$, $\pm100\,\mathrm{Hz}$), there are still some nodes where the BER persists at a high level.

\begin{figure}[!t]
\centering
\includegraphics[width=3in]{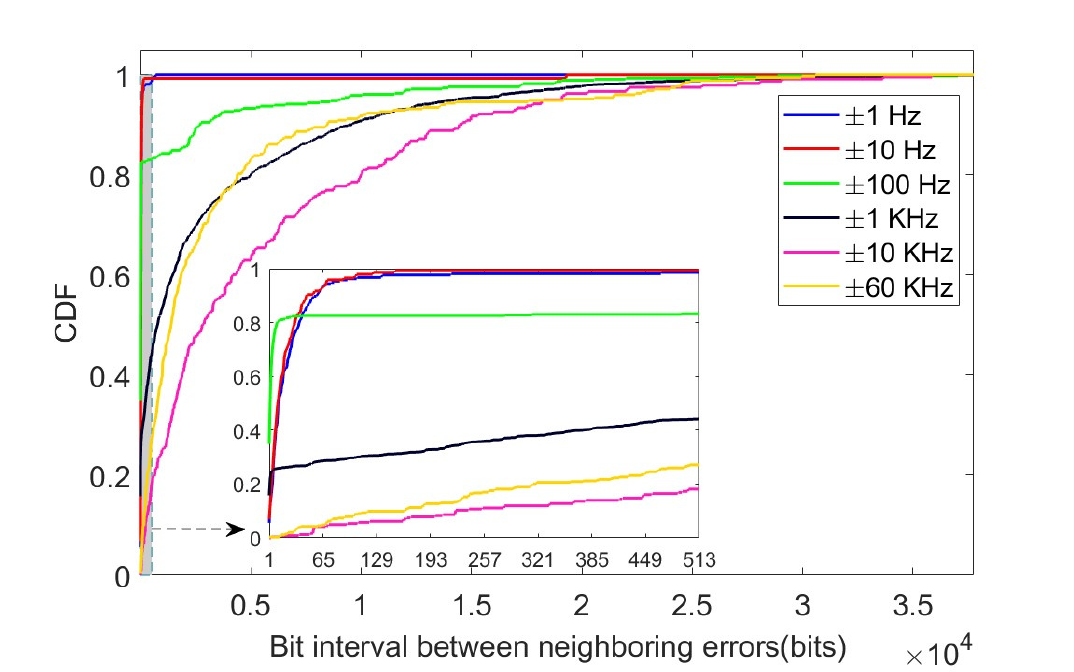}
\caption{The CDF of error spacing for different maximum CFO ranges}
\label{fig17}
\end{figure}

The impact of CFO variation on error distributions is further studied on the basis of the results of the above simulation. Specifically, we study the error spacing for every CFO variation range, which is defined as the interval (in number of bits) between subsequent bit errors. Fig.~\ref{fig17} shows the cumulative distribution function~(CDF) of error spacings, where we can observe that smaller CFO variation ranges are more likely to result in shorter error spacings (i.e., errors are clustered in bursts). Specifically, as Fig.~\ref{fig17} depicts, more than \SI{90}{\percent} of the error spacing is in the range from $1$~bit to $64$~bits, with the smaller CFO variation ranges of $\pm1\,\mathrm{Hz}$ and $\pm10\,\mathrm{Hz}$. When the CFO variation ranges larger than $\pm1\,\mathrm{KHz}$, less than \SI{10}{\percent} of the error spacings are shorter than $65$~bits. In other words, error bursts are more likely to occur when the CFO variation range is small compared to when it is large. The occurrence of error burst is attributed to the fact that smaller CFO fluctuations lead to a slower beating effect (i.e., alternations between constructive and destructive interference). As a result, the destructive interference tends to persist over a longer time period, covering multiple subsequent symbol periods, thereby causing clustered errors.

Interestingly, Fig.~\ref{fig17} also shows that at very large CFO variation ranges (e.g., $\pm60\,\mathrm{KHz}$ compared to $\pm10\,\mathrm{KHz}$), the errors become more bursty once again.
This phenomenon can be attributed to the short duration of destructive interference introduced by very large CFO values. When the CFO becomes sufficiently high, the destructive interference tends to have a much shorter temporal span, even shorter than a single symbol period, which means adjacent bit errors are more likely to fall into different beating periods, like the error spacing in CFO variation of $\pm60\,\mathrm{KHz}$. Under this condition, the influence of CFO variation on the sparsity of the error distribution becomes irregular and difficult to characterize. Moreover, larger error spacing in very large CFO variations, such as $\pm10\,\mathrm{KHz}$ and $\pm60\,\mathrm{KHz}$, also supports our hypothesis. As the error spacing becomes larger, it becomes increasingly likely that the neighboring errors are caused by distinct beating periods. 

\begin{figure*}[!t]
\centering
\subfloat[64-bit frames]{\includegraphics[width=2.3in]{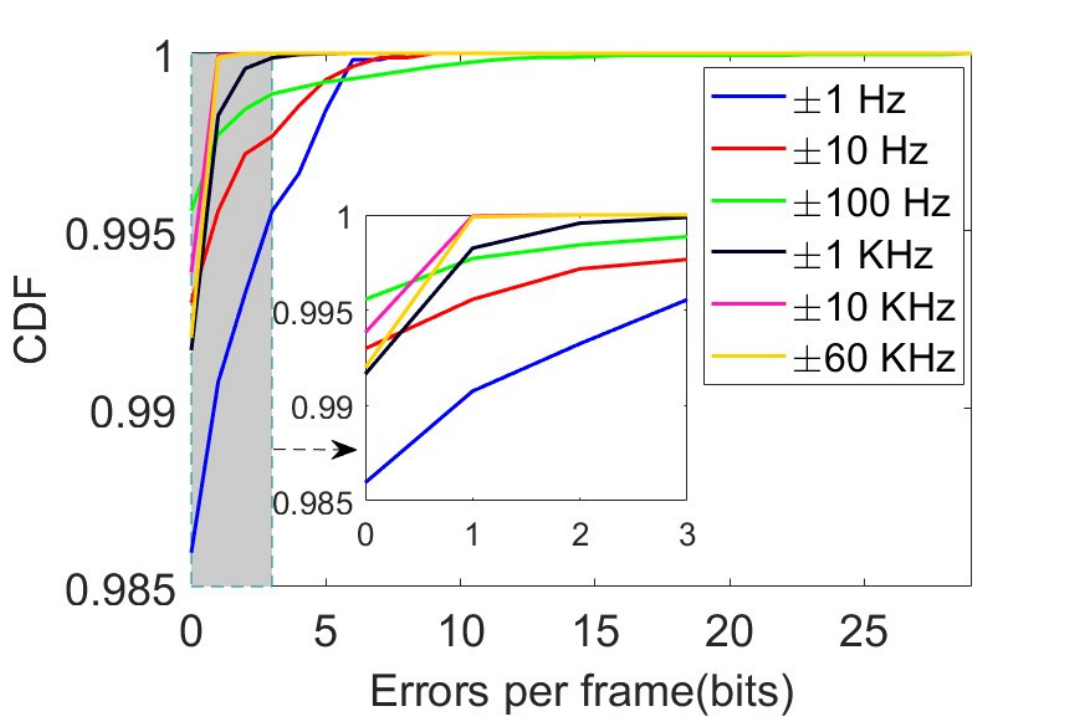}}
\label{64}
\hfil
\subfloat[128-bit frames]{\includegraphics[width=2.3in]{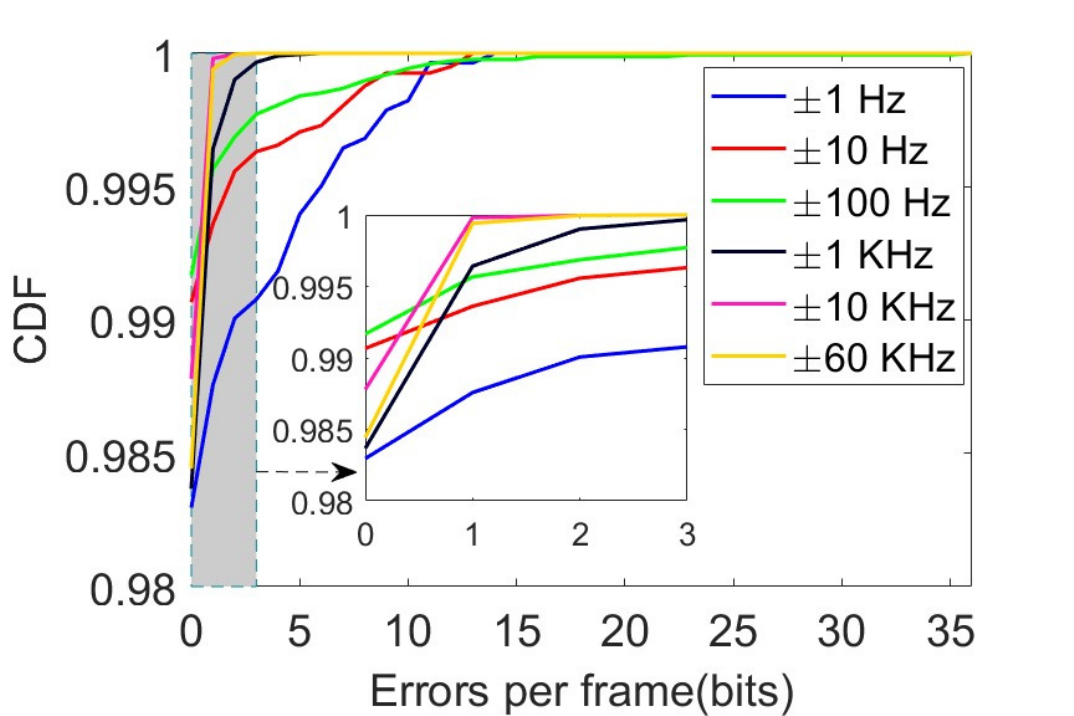}}
\label{128}
\hfil
\subfloat[256-bit frames]{\includegraphics[width=2.3in]{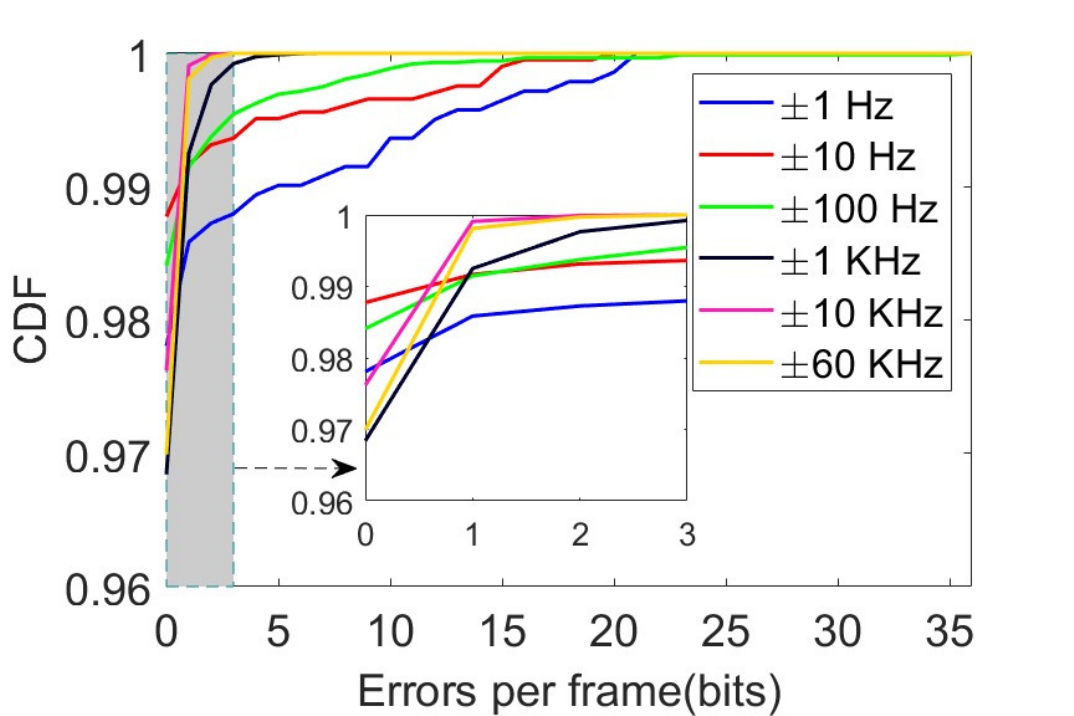}}
\label{256}
\caption{The CDF of the number of bit errors in frames of different sizes}
\label{fig18}
\end{figure*}

It is important to note that error clusters caused by small CFO variations pose particular challenges for error correction. Since most error correction codes are designed for sparse and \mbox{randomly-distributed} errors, the burst occurrence typically requires more complicated error-correction coding schemes or additional techniques (e.g., interleaving or Reed–Solomon codes). In order to further explore the effect of error bursts on \mbox{error-correcting} codes, we plot the number of bit errors per frame, considering different frame sizes from the set $\{64, 128, 256\}$ bits based on the same simulation results. 
In Fig.~\ref{fig18}, it can be seen that, for every size, almost all frames involve fewer than $3$ errors if the CFO variation range is larger than $\pm100\,\mathrm{Hz}$. In contrast, when the CFO variation range is below $\pm100\,\mathrm{Hz}$, a higher number of errors per frame is more likely for all frame sizes. 

\begin{figure}[!t]
\centering
\includegraphics[width=3in]{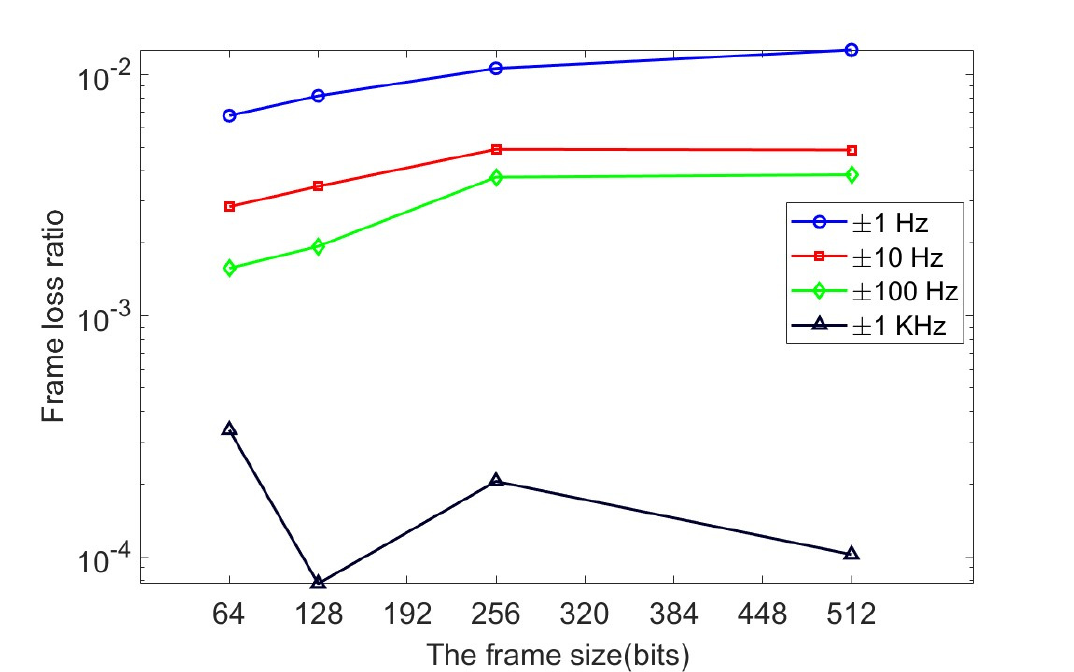}
\caption{The frame loss ratio as a function of the frame size for different CFO ranges. BCH code with code rate of \SI{80}{\percent} is applied}
\label{fig19}
\end{figure}

To showcase the degrading effect of bursty errors on error correction codes, we theoretically calculate the error correction capacity of the Bose-Chaudhuri-Hocquenghem~(BCH) code for different frame sizes. For a fair comparison, we fix the code rate at \SI{80}{\percent} regardless of the frame size. Assuming that a BCH code can successfully correct any number of bit errors within its correction capability, a frame is considered lost if the number of bit errors exceeds this capacity. Based on this assumption, the frame loss ratios under different CFO variation ranges are calculated and plotted as a function of frame sizes in Fig.~\ref{fig19}, and their standard deviations are listed in Table~\ref{table4}. As the figure depicts, the BCH codes perform worse under smaller CFO variation ranges, causing higher frame loss ratios. Specifically, BCH codes with \SI{80}{\percent} code rate are capable of limiting the frame loss ratios to around \SI{1}{\percent} under $\pm1\,\mathrm{Hz}$ CFO variation range, and to \SI{0.01}{\percent} under a $\pm1\,\mathrm{KHz}$ CFO variation range. For CFO variation range larger than $\pm1\,\mathrm{KHz}$, we did not observe any frame loss. It reveals that the system has the potential to achieve a frame loss ratio lower than \mbox{$7.11\times 10^{-6}$} in the transmission of $140625$ $64$-bit frames under a large CFO. 

\begin{table}[!t]
\caption{The standard deviations of frame loss ratio under various CFO ranges\label{table4}}
\centering
\begin{tabular}{c|c|c|c|c}
\hline
\textbf{CFO ranges} & \textbf{64-bit} & \textbf{128-bit} & \textbf{256-bit} & \textbf{512-bit}\\
\hline
$\pm1\,\mathrm{Hz}$ & \num{0.0079} & \num{0.0095} & \num{0.0127} & \num{0.0148}\\
$\pm10\,\mathrm{Hz}$ & \num{0.0044} & \num{0.0057} & \num{0.0072} & \num{0.0085}\\
$\pm100\,\mathrm{Hz}$ & \num{0.0013} & \num{0.0020} & \num{0.0039} & \num{0.0053}\\
$\pm1\,\mathrm{KHz}$ & \num{0.0006} & \num{0.0002} & \num{0.0005} & \num{0.0004}\\
\hline
\end{tabular}
\end{table}

The above results show that, in case CFO cannot be fully eliminated, a higher CFO range above $\pm1\,\mathrm{KHz}$ is desirable to limit burstiness of errors, and thus improve performance of error correction codes. In practical IoT scenarios, achieving perfect over-the-air carrier frequency synchronization is often infeasible, as nodes have limited power and resources, and hardware imperfections are common due to the use of cheap hardware. In addition, approaches for CFO estimation and compensation are not easily applicable to the proposed protocol, since additional latency is potentially introduced, and small CFOs, caused by low-accuracy CFO estimation, can degrade the system reliability. As a result, instead of eliminating the CFOs, a potential (counter-intuitive) solution could be to add a random CFO (e.g., higher than $\pm1\,\mathrm{KHz}$) to every transceiver, in order to avoid error clustering resulting from small CFOs. Note that the optimal value of this random CFO depends on the values of $T_p$ and $T_s$ selected for RF-Zero-Wire communication.

\subsection{Network scalability evaluation}
We evaluate the network scalability by varying the grid distance $d$, the node density, the data rate, and the channel profile. To evaluate the effect of grid distance $d$ on the network reliability, a data rate of $40$~Kbps is used in a \mbox{$16$-node} network~($4$-by-$4$), and the grid distance $d$ is varied from $5$~m to $11$~m. Similarly to prior experiments, we change the CFOs in a certain range every second and transmit $600$K bits for every $d$. The average BER across all nodes is used as a metric. 

\begin{figure*}[ht]
    \centering
    \begin{minipage}[t]{0.32\textwidth}
        \centering
        \includegraphics[width=\linewidth]{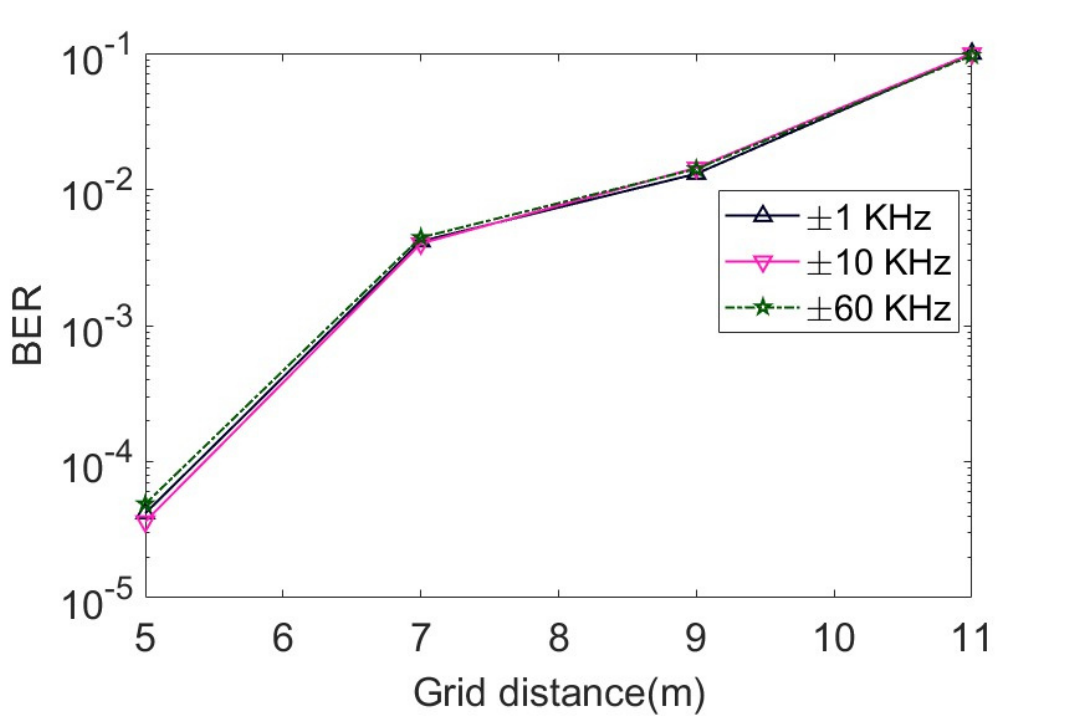}
        \caption{The effect of grid distance on network reliability}
        \label{fig20}
    \end{minipage}
    \hfill
    \begin{minipage}[t]{0.32\textwidth}
        \centering
        \includegraphics[width=\linewidth]{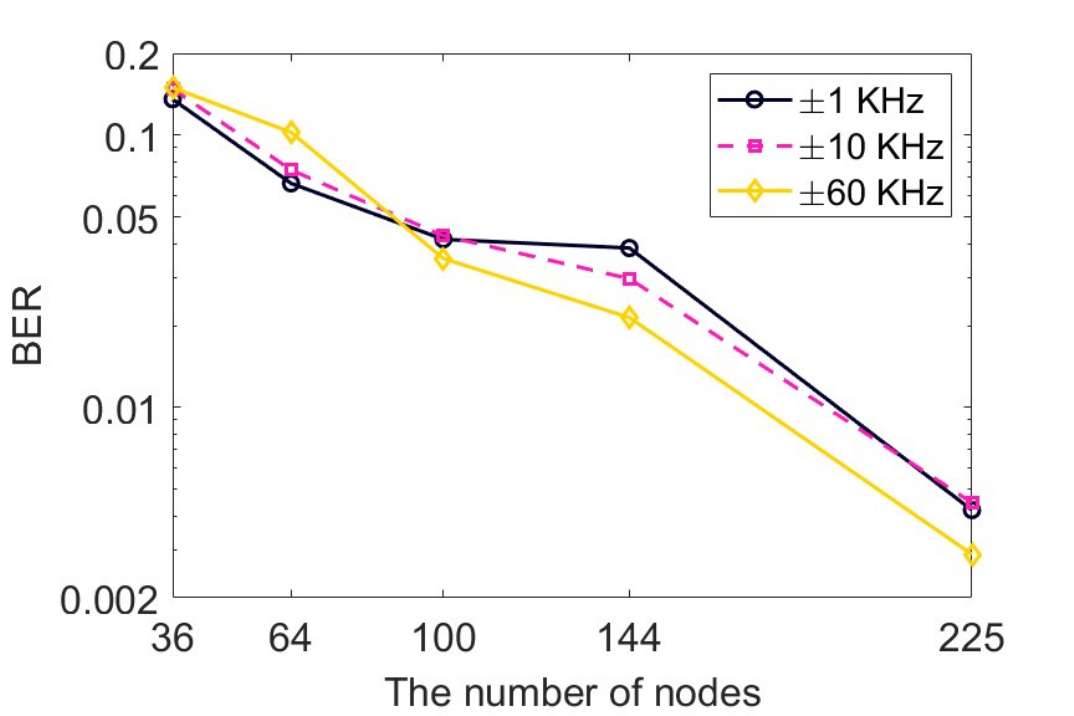}
        \caption{The effect of node density on network reliability}
        \label{fig21}
    \end{minipage}
    \hfill
    \begin{minipage}[t]{0.32\textwidth}
        \centering
        \includegraphics[width=\linewidth]{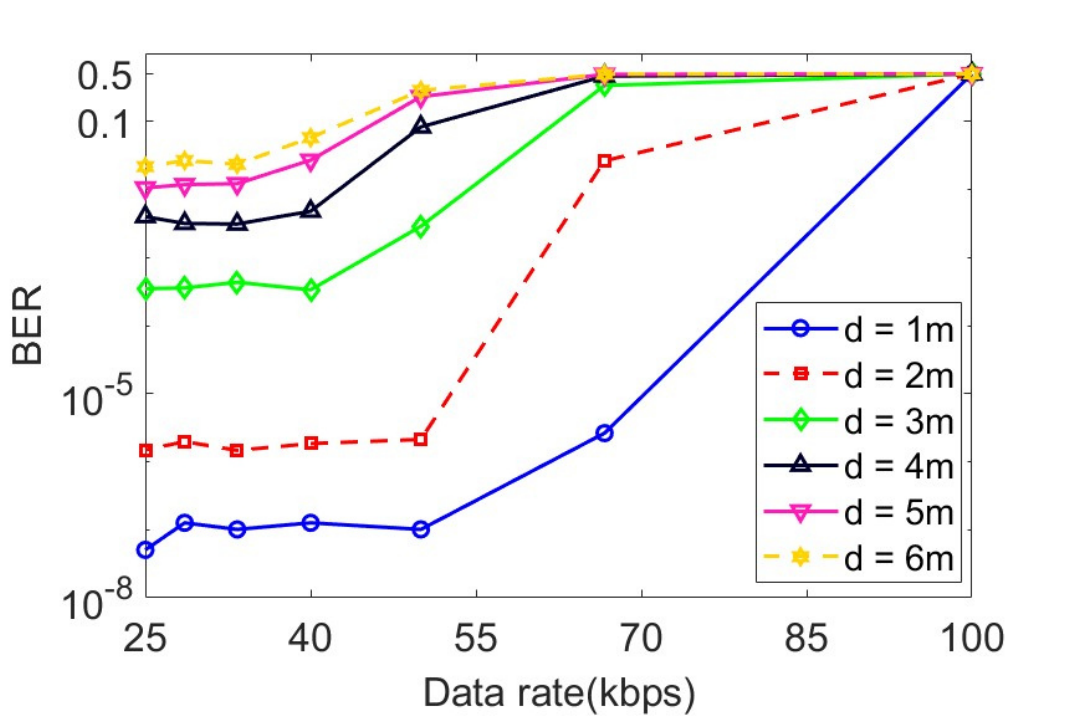}
        \caption{The effect of data rate on network reliability}
        \label{fig22}
    \end{minipage}
\end{figure*}

Fig.~\ref{fig20} illustrates the effect of grid distance on network BER under different CFO variation ranges. It shows that the network reliability dramatically increases with the grid distance. For the short grid distance of \SI{5}{\meter}, a small BER less than \SI{0.01}{\percent} can be achieved, while the longer grid distance of \SI{11}{\meter} introduces BER of about \SI{10}{\percent}. This increase occurs because with the increase of $d$, fewer nodes are located in the transmission range of each hop. Subsequently, there are fewer nodes cooperatively forwarding the information per hop, thereby reducing the received signal strength. In addition, the longer grid distance inherently extends the transmission distance between relay nodes, thus decreasing the received signal strength at each receiver. Furthermore, Fig.~\ref{fig20} confirms our conclusion that the effect of various CFO variation ranges on BER is negligible compared to the grid distance. 

We also explore the effect of node density on BER. In a fixed area of \SI{3600}{\square\meter}, we vary the node density by placing different numbers of nodes, such as $36$~($6$-by-$6$), $64$~($8$-by-$8$), $100$~($10$-by-$10$), $144$~($12$-by-$12$), and $225$~($15$-by-$15$)~nodes, which are arranged in a grid layout with different grid distance. Here, the corresponding grid distances are \SI{10}{\meter}, \SI{7.5}{\meter}, \SI{6}{\meter}, \SI{5}{\meter}, and \SI{4}{\meter}. The data rate is fixed to $40$~Kbps, and CFOs change every second within the specified range. As Fig.~\ref{fig21} shows, in a fixed area, higher density helps to improve network reliability, compared to a sparse topology. Specifically, when the number of nodes increases from $36$ to $225$, the BER is reduced from \SI{16.5}{\percent} to \SI{0.3}{\percent}, which contrasts with conventional wireless networks, where more nodes degrade the reliability due to increased contention, collisions, and back-offs. The behavior makes the proposed protocol particularly suitable for dense IoT deployments, as reliability improves when the network becomes denser. This improvement is attributed to more concurrent transmissions enabled by the higher node density. At the same time, the higher node density results in shorter transmission distance between neighbouring relays, resulting in higher amplitude of received signals.

Then, we explore the effect of the data rate on the network reliability. In this experiment, we independently vary the data rate in a $100$-node network with grid distances ranging from \SI{1}{\meter} to \SI{6}{\meter}. The CFO variation range is fixed to $\pm10\,\mathrm{KHz}$ and the CFO changes randomly every second. The choice of $100$ nodes ensures that the network includes a sufficient range of hop counts, typically between $2$ and $7$. The results in Fig.~\ref{fig22} are averaged over a duration of \SI{15}{\second}. Fig.~\ref{fig22} reveals that a higher data rate potentially reduces the network reliability. As shown in Fig.~\ref{fig22}, all the curves, representing the BER versus data rate under different grid distances, follow a similar trend: initially, the curve is mostly flat at lower data rates, followed by a dramatic increase, then gradually approaching \SI{50}{\percent}.

The root cause for the increase of BER at higher data rates can be attributed to the presence of ISI, as illustrated in Fig.~\ref{fig5}. As mentioned in Section~\ref{sec3_3}, the symbol period should be long enough to synchronize the detection and relaying of symbols by nodes within the mutual communication range, which avoids the echoes of the previous symbol interfering with the detection of the next symbol. However, with the increase in data rate, the symbol period becomes too short, leading to ISI. The ISI will seriously affect the detection of $0$ symbols succeeded or preceded by $1$ symbols. If the echoes of a $1$-symbol arrive in the time slot of a $0$-symbol, the $0$-symbol will be wrongly detected as $1$. Most importantly, the wrong detection of the symbol $0$ will trigger the node to incorrectly relay a $1$-symbol, propagating the error to more nodes. In contrast, ISI does not affect the detection of $1$-symbols in our protocol, as a $0$-symbol is represented by a silent period. Fig.~\ref{fig22} shows that ISI-induced degradation in BER becomes more severe as the data rate increases. The magnitude of the ISI-induced degradation depends on the strength of the echoes. Shorter symbol periods~(i.e., higher data rate) allow echoes from more nodes, increasing the strength of echoes and furthering ISI impact. In the worst case, the ISI causes an approximate \SI{50}{\percent} BER at the data rate of $100$~kbps (i.e., nearly all $0$ symbols are incorrectly detected as $1$). 

The figure also shows that the maximum achievable data rate (i.e., the point where the BER curve suddenly increases to nearly 50\%) is inversely related to the distance between neighboring nodes. For example, for the grid distance of \SI{1}{\meter}, a sharp increase is present when the data rate is higher than $66.7$~kbps. However, this shift occurs already at $50$~kbps and $40$~kbps for a grid distance of \SI{2}{\meter} and \SI{3}{\meter}, respectively. As the transmit power and number of nodes remain constant, a lower distance $d$ results in a more compact network, where symbols propagate faster (thus lower hop count). This also means that the time interval over which echoes of a symbol may be heard by any node is reduced, thus resulting in less chance of ISI at lower symbol durations $T_s$. As the data rate is inversely proportional to $T_s$, a more compact network (i.e., with lower $d$) thus supports a higher data rate.

In other words, Fig.~\ref{fig22} reveals that the network bottleneck is determined by the network diameter, which may be affected by the number of nodes or the grid distance. Specifically, if we fix the grid distance, the network diameter will increase as the number of nodes increases. Conversely, increasing the grid distance with a fixed number of nodes will also increase the network diameter. Once the network diameter exceeds the maximum hop count that the data rate can support, expressed as $\frac{T_s}{r}$, ISI may emerge, thus degrading network performance. In other words, the data rate imposes a bottleneck on network diameter that can be reliably supported under symbol-synchronous transmission.

\begin{figure}[!t]
\centering
\includegraphics[width=3in]{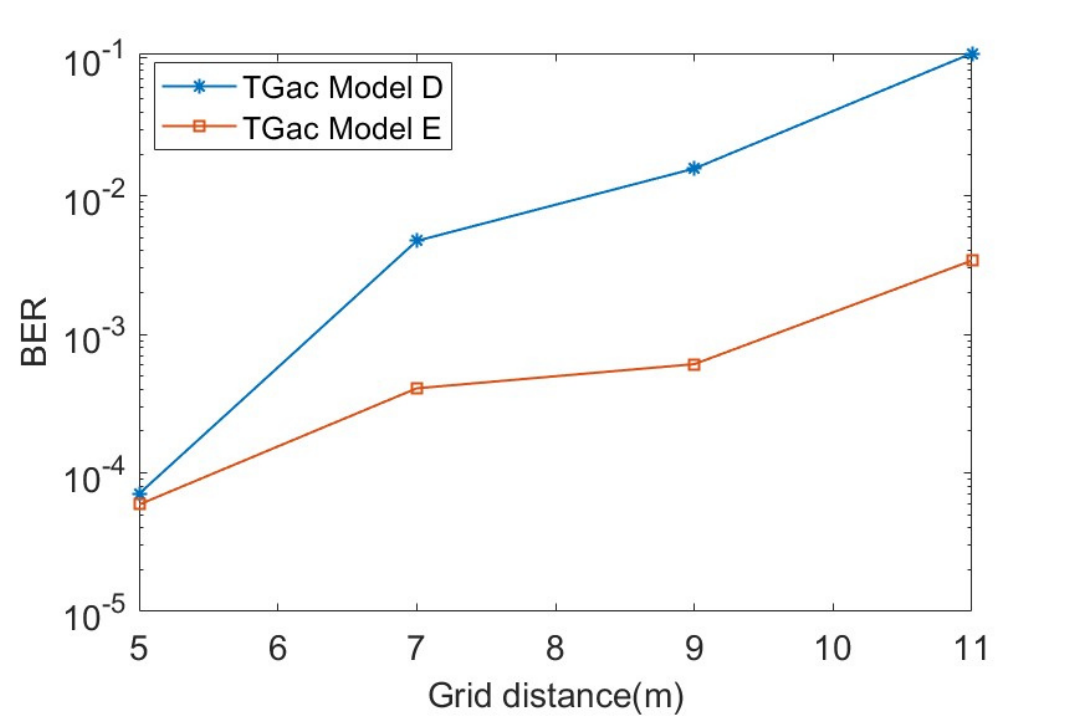}
\caption{The effect of channel conditions on network reliability}
\label{fig23}
\end{figure}

Finally, we evaluate the effect of channel conditions in different environments on the network performance. The network reliability on various grid distances under the TGac channel model E is evaluated to compare with TGac channel model D. The transmission with a $40$~Kbps data rate and $\pm10\,\mathrm{KHz}$ CFO range in a $16$-node network ($4$-by-$4$) is simulated.

As Fig.~\ref{fig23} shows, the effect of the grid distance on the network reliability shows the same trend under different propagation environments. However, the network reliability in the Model E is higher than that in the Model D, since Model E models a more open indoor environment, such as a gymnasium or large hall, and has a larger Rician K-factor. A higher K-factor indicates a stronger line-of-sight component, thus leading to a higher received signal power in the same transmission distance.
\subsection{Evaluation of the window-length-based trade-off}
As mentioned in Section~\ref{sec3_4}, the detection delay, the time from the node wake-up to the successful detection of a symbol, and the SER are sensitive to the window length. In this section, we quantify how the window length balances the detection accuracy of symbols $1$ and $0$, and affects the detection delay. We vary the window length from $5$ to \SI{20}{\micro\second} based on a typical network configuration, where the data rate, grid distance, the number of nodes, and the CFO range are respectively set to $40$ Kbps, $5$ meters, 25, and $\pm10\,\mathrm{KHz}$. In total, $600000$ bits are transmitted for robust evaluation.

\begin{figure}[!t]
\centering

\subfloat[Detection latency]{%
    \includegraphics[width=3in]{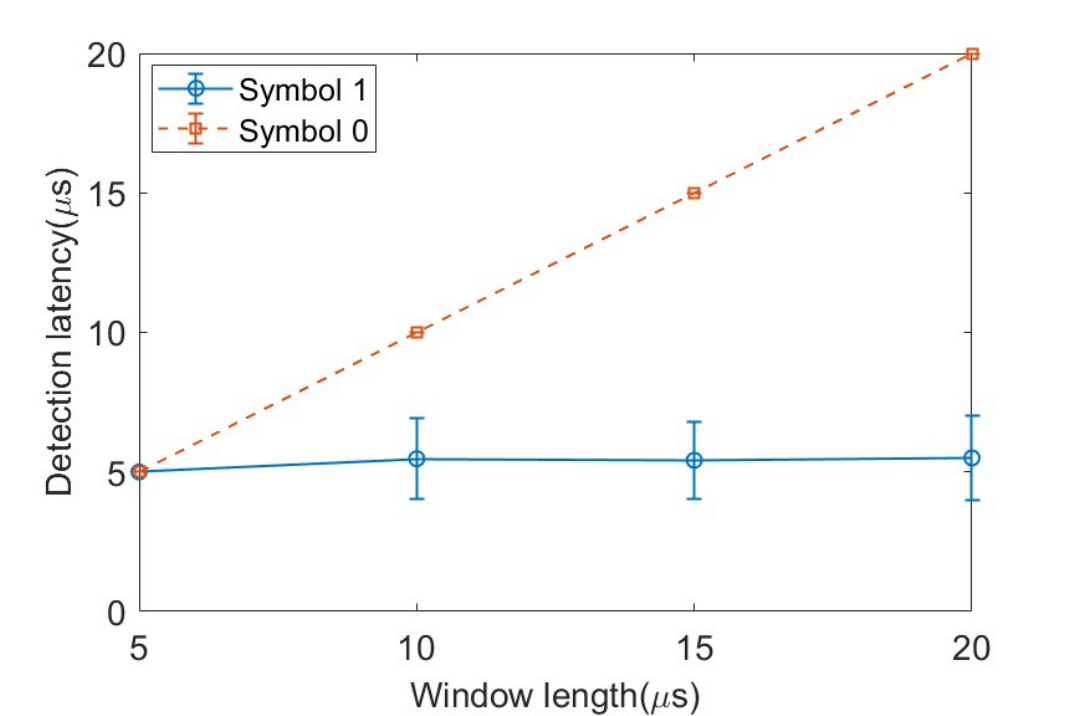}%
    \label{fig24_1}%
}
\hfil
\subfloat[SER]{%
    \includegraphics[width=3in]{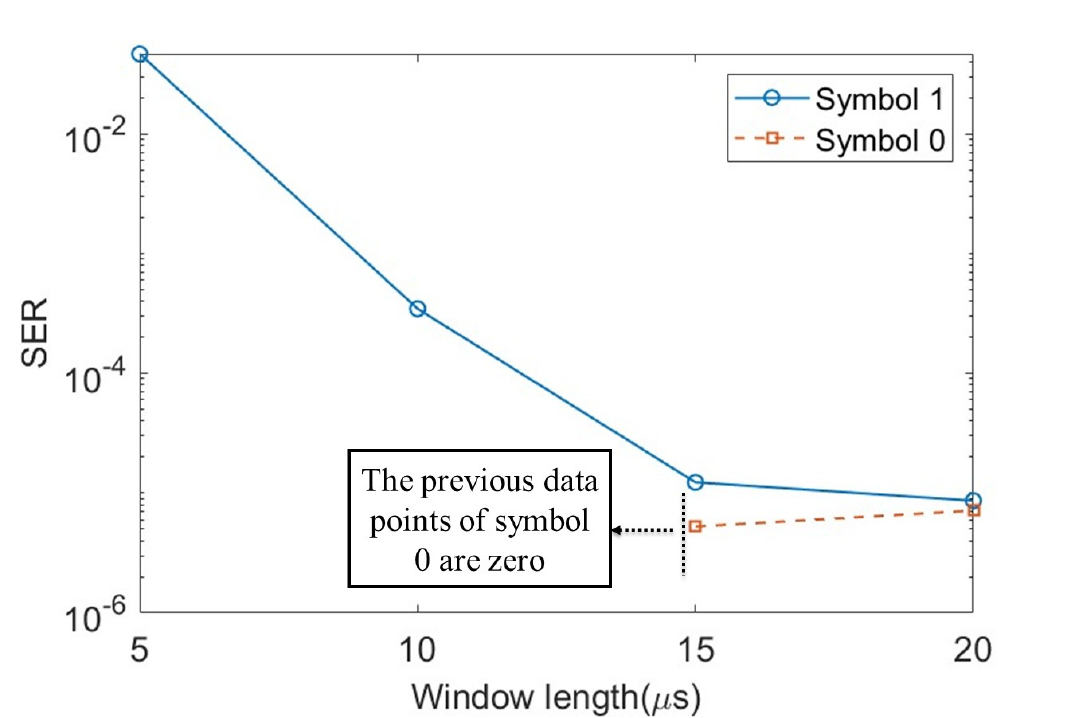}%
    \label{fig24_2}%
}

\caption{The effect of the window length on detection latency and accuracy}
\label{fig24}
\end{figure}

Fig.~\ref{fig24_1}  shows the average detection delay as a function of the window length. As shown, the detection latency increases linearly with the window size for symbol $0$, while it remains nearly constant for symbol $1$. This is due to the fact that the detection of symbol $0$ requires the node to keep observing during the whole window, while symbol $1$ is still likely to be detected at the first attempt in a longer window. Moreover, a trade-off between the SER   of symbol $1$ and symbol $0$ is depicted in Fig.~\ref{fig24_2}. The result complies with our theoretical analysis in Section~\ref{sec3_4}. A longer window length increases the SER of symbol $0$, while decreasing the SER of symbol $1$.

In this article, we choose \SI{10}{\micro\second} as the window length, which provides a well-balanced compromise between detection accuracy and latency. As Fig.~\ref{fig24_2} illustrates, with the increase of window length from $5$ to \SI{10}{\micro\second}, the SER of symbol $1$ shows a significant decrease. At the same time, no increase in SER for symbol $0$ could be detected. If the window length further increases, the improvement of the SER of symbol $1$ is limited, while introducing additional SER for symbol $0$, as well as increasing the detection delay.

\section{Disscusion}
\subsection{Practical considerations}
This section discusses practical considerations related to hardware impairments, such as oscillator imperfections and processing delays, and analyzes their effect on the performance of the proposed protocol.

The oscillator quality directly affects the CFO range. Specifically, a high-quality oscillator like the one used in the USRP N310 SDR, with a $0.1$~ppm deviation~\cite{ettus2019n310}, results in a small CFO range of $\pm250\,\mathrm{Hz}$ on the carrier band of \SI{2491}{\mega\hertz}. A low-quality oscillator like the one used in ADALM-PLUTO SDR, with a $25$~ppm deviation~\cite{rtlsdr2017pluto}, results in a large CFO range of $\pm60\,\mathrm{KHz}$. Therefore, in this article, different levels of CFO that can exist in various hardware platforms were simulated and evaluated. Simulation results show that a large CFO is more beneficial for error correction. It means that the system can operate on low-quality hardware with an imperfect oscillator. On high-quality hardware, an additional “artificial” CFO can be introduced on purpose to increase the CFO range. This shows the robustness of our proposed method, independent of oscillator quality, in terms of CFO.

In addition to the CFO range, oscillator imperfections can also induce some deviations in the detection window positions across symbol periods. However, in this article, the window length is set to be over three times the pulse duration, which means the pulse can still be captured and detected as long as the offset is within the window range. At the same time, when the window position is initially configured, a time offset is applied according to Equation~\eqref{eq3} to compensate for clock drift and inaccuracy, thereby reducing the impact of oscillator imperfections on window accuracy.

Moreover, in a practical implementation, variations in the processing delay would affect the relay time per hop, thereby increasing the latency. At the same time, a longer symbol period (lower data rate) is needed to support the same number of hops with a longer relay time. In this article, it is assumed that the processing delay is small enough to be negligible. The real processing delay would significantly depend on the implementation details (e.g., in FPGA, bare metal, Linux OS). As such, it is hard to select a representative processing delay value for use in simulations, but an implementation of RF Zero-Wire on an FPGA is planned in the future, which can be used to derive representative values.

\subsection{Implications for energy efficiency}
Benefitting from concurrent transmissions and symbol-level flooding, the proposed protocol is also energy-efficient and could be an effective alternative to short-range low-power protocols, such as Bluetooth and IEEE 802.15.4. Thanks to the use of concurrent transmissions, the additional power consumption resulting from routing and channel access is avoided. Thanks to symbol-level flooding, the tight time synchronization needed in packet-level flooding, which is hard and costly to achieve in energy-constrained devices, is avoided. Moreover, since symbol-synchronous transmission relies on constructive interference among concurrent transmissions of multiple relays for signal detection, the sink node can still correctly detect the transmitted symbol even if only a subset of relays forwards each symbol, provided that the received signal strength is sufficiently high to be distinguished from noise. This property enables symbol-synchronous communication to naturally support adaptive relay duty cycles, where each low-power relay forwards only a subset of symbols according to its available energy. As a result, the overall energy consumption of the network can be dynamically adjusted without compromising the reliability of symbol delivery. This will be further studied in future work.

\subsection{Distance consideration}
In this article, to maintain alignment with the hardware experiments, a maximum grid distance of \SI{11}{\meter} is considered in the simulations. This distance is constrained by the chosen transmit power of $0$~dBm, the noise floor of -$60$~dBm, and the propagation environment, not by the protocol design. If the transmitter power is increased or the noise floor is lower, the maximum single-hop communication distance increases accordingly.

\section{Conclusion and outlook}
\label{sec6}
This article introduces a novel communication protocol based on RF-based symbol-synchronous transmission to reduce the end-to-end latency of wireless multi-hop networks. It challenges the traditional network protocols comprising collision-free routing, store-and-forward switching, and medium access control. In contrast, RF-Zero-Wire utilizes concurrent transmissions and enables the network to disseminate the information symbol-by-symbol instead of waiting for the complete frame reception. We used a testbed consisting of $4$ SDRs to study the effects of constructive and destructive interference and select parameters for simulation. Specifically, we analyze how the CFO, inevitably introduced by hardware imperfections, affects the reliability of concurrent transmissions. System-level simulations in MATLAB demonstrate that RF-Zero-Wire maintains a delay of sub-milliseconds for small frame transmissions~(below $4$ bytes) over $5$~hops. For a larger, $128$-byte frame, the end-to-end latency is around $25$~ms. Moreover, the latency increase of around \SI{5}{\micro\second} per extra hop remains consistent in the transmission of frames of arbitrary sizes, which is only about \SI{0.16}{\percent} of the end-to-end network latency. Additionally, through the simulation of different CFOs, this article reveals a counterintuitive phenomenon that smaller CFOs are more likely to cause error bursts, degrading the network reliability when making use of error-correcting codes. The network scalability evaluations demonstrate that shorter grid distances and denser node deployments help to improve network reliability. We also investigate the effect of the selected data rate on the network reliability for different grid distances.  

Future research may focus on the energy efficiency of the protocol, particularly in dense networks such as robotic swarms or WSNs, where power consumption becomes a critical concern. We also consider differentiating the power of concurrent transmitters to mitigate destructive interference. At the same time, a proper power management strategy will be developed when applying the power differentiation strategy. Another promising direction is the adoption of more sophisticated modulation schemes to improve transmission reliability and data rate. When designing the modulation scheme for a symbol-synchronous protocol, a significant challenge is that the modulated information should be robust to the signal superimpositions in concurrent transmission. And the corresponding demodulation should have low complexity to be performed with low latency. Therefore, Gaussian frequency shift keying~(GFSK) and Manchester coding could be considered in future work. Additionally, security aspects and robustness against (malicious) interference remain an open challenge. And the node mobility will be taken into account in the channel model and protocol design for future work.

\bibliographystyle{IEEEtran}
\bibliography{reference}

\begin{IEEEbiography}[{\includegraphics[width=1in,height=1.25in,clip,keepaspectratio]{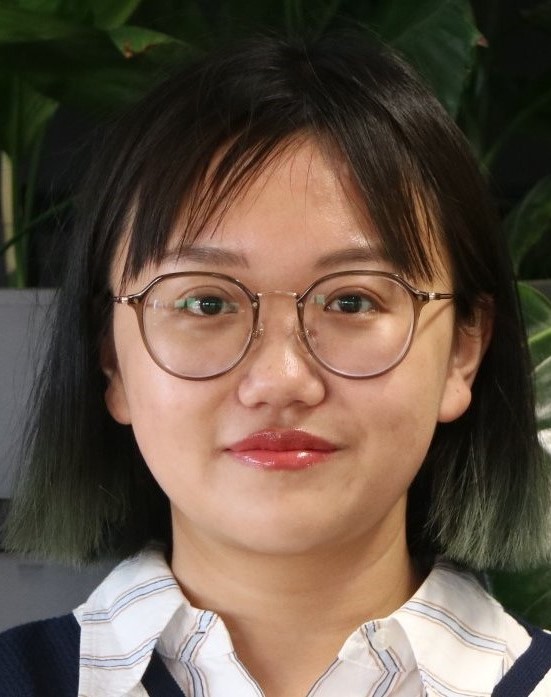}}]{Xinlei Liu}
received the B.Sc. degree in electronic information engineering from Nanjing University of Information Science and Technology, China, in 2020, and the M.Sc. degree in engineering of digital systems and telecommunications from the Australian National University, Australia, in 2022. She is currently pursuing her Ph.D. in the IDLab research group, University of Antwerp, and imec, Belgium. Her research interests include design and modeling of the PHY layer in wireless communication, wireless network protocols, and embedded software. 
\end{IEEEbiography}
\begin{IEEEbiography}
[{\includegraphics[width=1in,height=1.25in, clip,keepaspectratio]{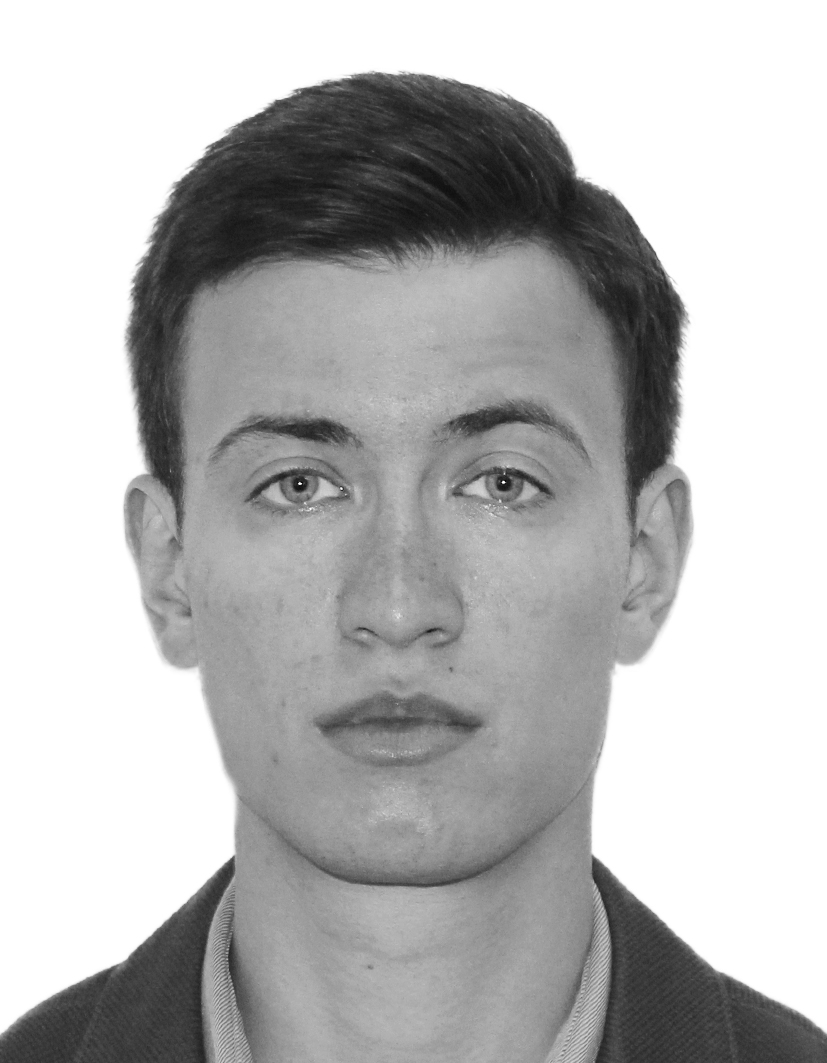}}]{Andrey Belogaev} received his M.Sc. and Ph.D. degrees in telecommunications from Moscow Institute of Physics and Technology in 2016 and 2020, respectively. Subsequently, he worked as a senior researcher at the HSE University and IITP RAS until 2022. Currently, he is a senior researcher in the IDLab research group at the University of Antwerp and imec, Belgium. His research interests include design and optimization of protocols in wireless systems for high-performance and low-power communication.
\end{IEEEbiography}
\begin{IEEEbiography}[{\includegraphics[width=1in,height=1.25in,clip,keepaspectratio]{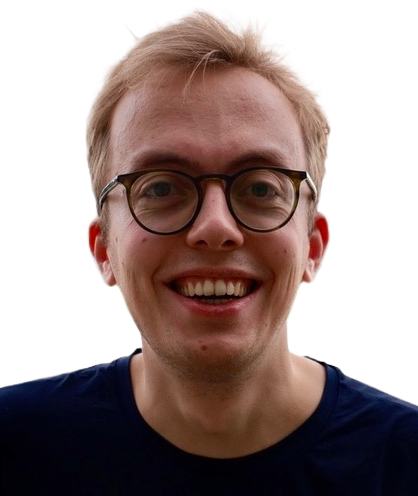}}]{Jonathan Oostvogels} is a postdoctoral researcher at KU Leuven, Belgium, where he also obtained his Ph.D. in 2024. Previously, he was a visiting research lead with the AIO team at Inria Paris, France. His research interests include the design of low-latency and densely deployed embedded systems. His work was recognised as a Communications of the ACM Research Highlight, received a Best Paper Award at ACM SenSys, and received a Best Presentation Award at ACM MobiSys’ Rising Star Forum.
\end{IEEEbiography}
\begin{IEEEbiography}[{\includegraphics[width=1in,height=1.25in,clip,keepaspectratio]{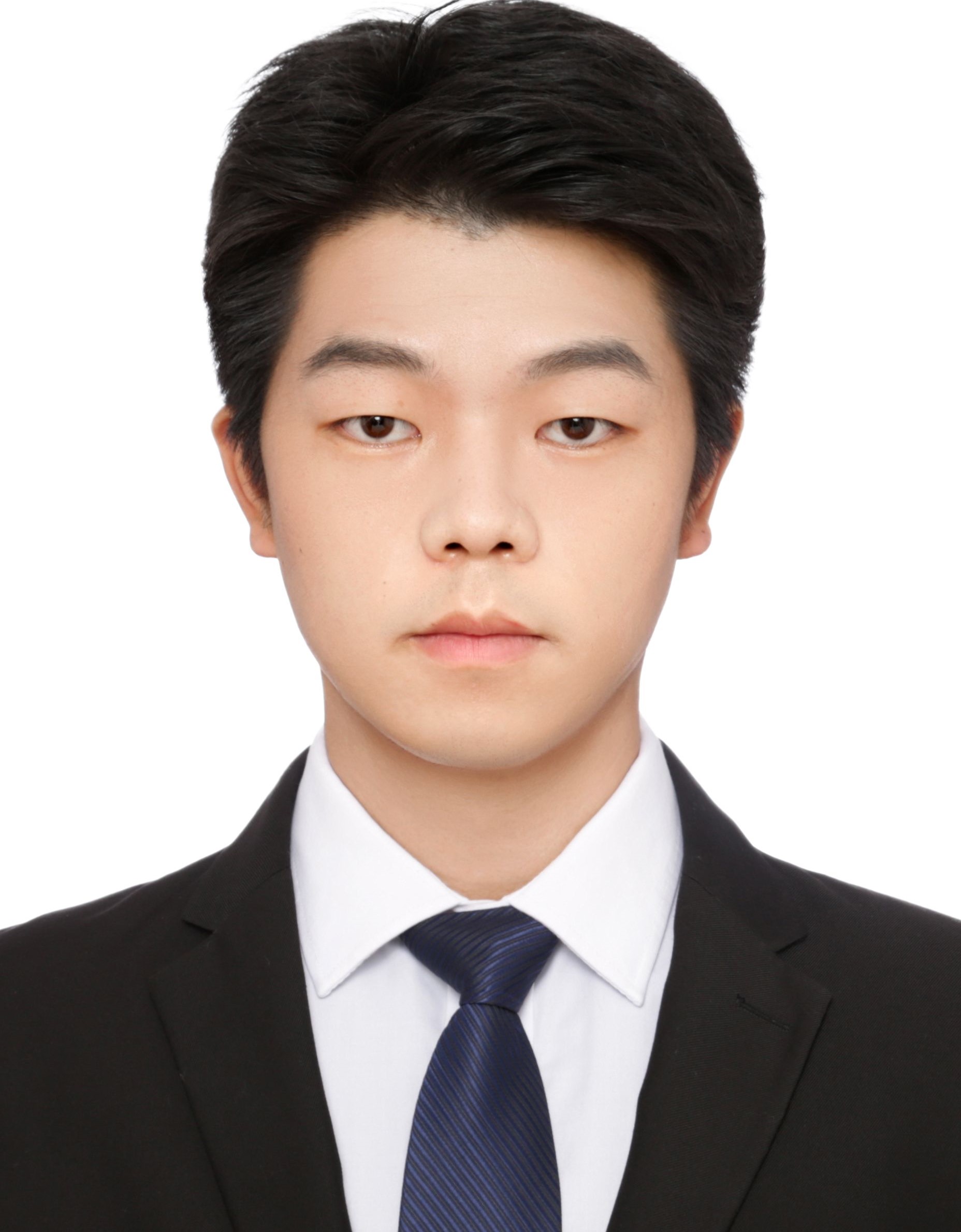}}]{Bingwu Fang}
received the B.Sc. degree in electronic and information engineering from Beijing Jiaotong University, Beijing, China, in 2018, and the M.S. degree in electronic and information engineering from Beihang University, Beijing, China, in 2021. He is currently pursuing the Ph.D. degree in computer science at KU Leuven, Belgium. His research interests include low-latency wireless communications and networked systems.
\end{IEEEbiography}
\begin{IEEEbiography}[{\includegraphics[width=1in,height=1.25in,clip,keepaspectratio]{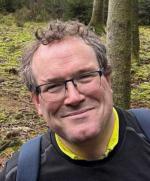}}]{Danny Hughes}
is a Professor with the Department of Computer Science of KU Leuven (Belgium), where he is a member of the DistriNet research group and leads the Networked Embedded Software taskforce. Danny has a PhD from Lancaster University (UK) and has since worked as a Visiting Scholar with the University of California at Berkeley (USA), a Visiting Scholar with the University of Sao Paulo (Brazil) and as a Lecturer with Xi'an Jiaotong-Liverpool University (China). His PhD focused on Peer-to-Peer (P2P) systems and his current research is on distributed software systems and the Internet of Things (IoT).
\end{IEEEbiography}
\begin{IEEEbiography}[{\includegraphics[width=1in,height=1.25in,clip,keepaspectratio]{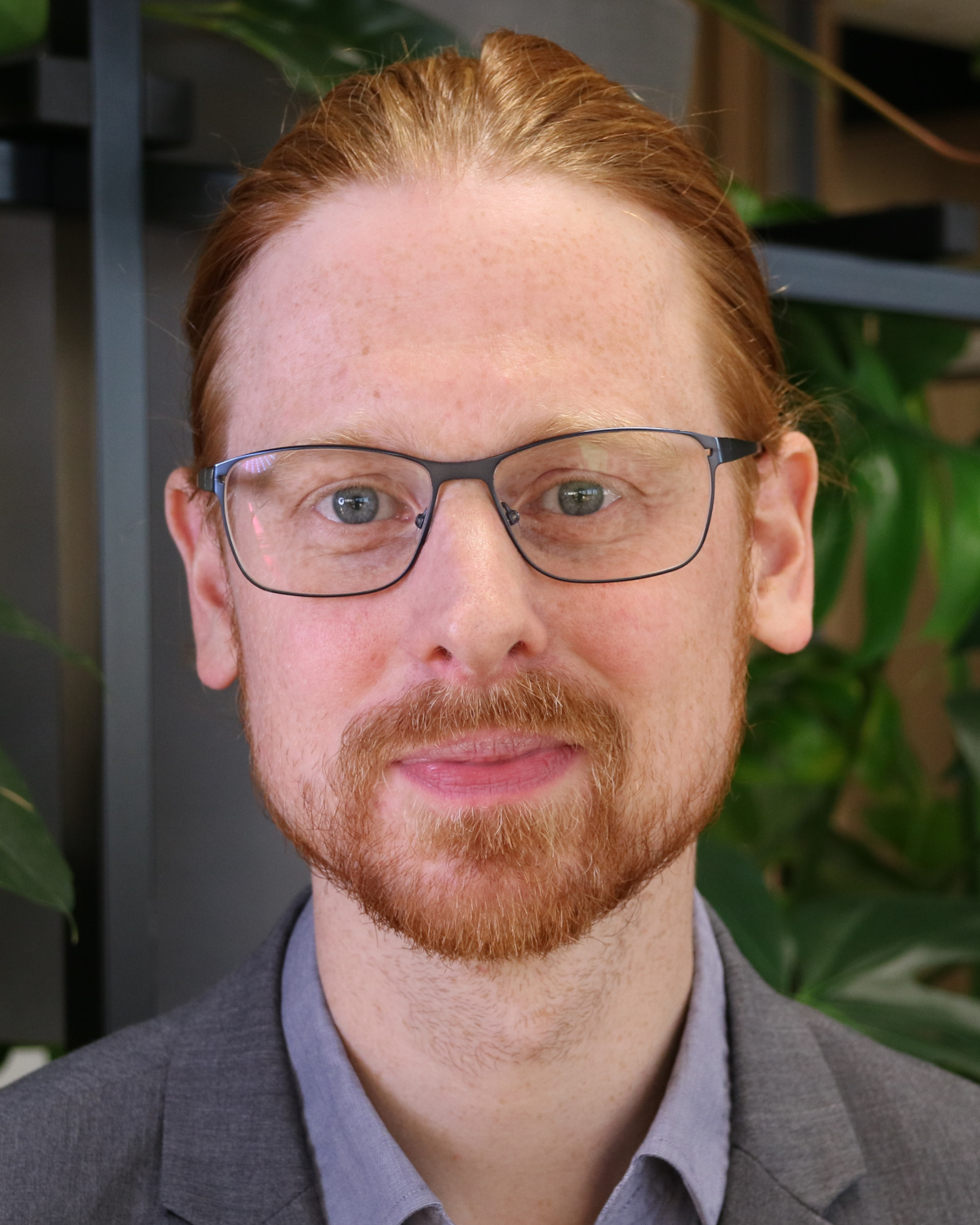}}]{Jeroen Famaey}
is an associate professor at the University of Antwerp, Belgium, and a senior researcher at imec, Belgium. He leads the Perceptive Radio Systems team in IDLab research group, performing research on wireless communications and sensing. He received his M.Sc. and Ph.D. in Computer Science Engineering from Ghent University, Belgium, in 2007 and 2012, respectively. His current research interests include low-power distributed machine learning and wireless communications, as well as data-driven integrated sensing and communications. He has co-authored over 200 papers, published in international peer-reviewed journals and conference proceedings.
\end{IEEEbiography}

\vspace{11pt}

\vfill

\end{document}